\documentclass[fleqn,usenatbib]{mnras}

\usepackage{afterpage}
\usepackage{newtxtext,newtxmath}
\usepackage{array}
\usepackage{multirow}
\usepackage{gensymb}
\usepackage{makecell}
\usepackage[T1]{fontenc}

\DeclareRobustCommand{\VAN}[3]{#2}
\let\VANthebibliography\thebibliography
\def\thebibliography{\DeclareRobustCommand{\VAN}[3]{##3}\VANthebibliography}

\usepackage{graphicx}	% Including figure files
\usepackage{amsmath}	% Advanced maths commands
\usepackage[final]{changes}

\usepackage{natbib} % For citation commands required by mnras

\title[Prominence Analysis on Lyman Lines]{Analysis of Lyman $\beta$ and Lyman $\gamma$ Lines in a Pre-Eruptive and Eruptive Prominence with Solar Orbiter SPICE Observations
}

\author[Y. Zhang et al.]{
Yong Zhang,$^{1}$
Nicolas Labrosse,$^{1}$
Susanna Parenti$^{2}$
and Therese A. Kucera$^{3}$
\\
$^{1}$SUPA School of Physics and Astronomy, University of Glasgow, Glasgow G12 8QQ UK\\
$^{2}$Institut d'Astrophysique Spatiale, Bat 121, Université Paris-Saclay/CNRS 91405 Orsay Cedex France\\
$^{3}$Heliophysics Science Division, NASA Goddard Space Flight Center, Greenbelt, MD, 20771 USA%\\
}

\date{Accepted XXX. Received YYY; in original form ZZZ}

\pubyear{2015}

\begin{document}
\label{firstpage}
\pagerange{\pageref{firstpage}--\pageref{lastpage}}
\maketitle

% Abstract of the paper
\begin{abstract}
The first dedicated observation of an off-limb prominence by the Spectral Imaging of the Coronal Environment (SPICE) instrument on board Solar Orbiter took place on April 15, 2023. Our aim is to provide an overview of the potentiality of the diagnostics using these data. We show that we can derive the changes in the physical parameters of the pre-eruptive and eruptive prominence using the Lyman lines. We investigate the integrated intensity and line widths of the Lyman $\beta$ and Lyman $\gamma$ lines, finding  variations between the prominence, disk, and coronal regions. The results reflect dynamic changes in density, temperature, and optical thickness. We analyze the spatial and temporal evolution of the Lyman $\beta$ and Lyman $\gamma$ line profiles. Using a simple geometric model, we obtain the radial velocity of this prominence at the early phase of its eruption with GONG H$\alpha$ images. This offers a way of calculating the radial velocity of an eruptive filament from a pair of 2D images. The result helps us understand the potential Doppler effect in line profiles. Overall, the spectral profiles indicate that the eruption enhances spatial and temporal variations in line intensity, reflecting dynamic changes in plasma conditions within the prominence. These findings highlight the diagnostic potential of SPICE observations, and future Non-LTE radiative transfer modeling will help to further constrain prominence plasma parameters.   
\end{abstract}


%%%%%%%%%%%%%%%%%%%%%%%%%%%%%%%%%%%%%%%%%%%%%%%%%%

%%%%%%%%%%%%%%%%% BODY OF PAPER %%%%%%%%%%%%%%%%%%

\section{Introduction}

% The prominence was also observed by the Solar Terrestrial Relations Observatory Ahead (STEREO-A) and GONG H$\alpha$ telescope of El Teide. 

 The first dedicated observation of an off-limb prominence by Solar Orbiter took place on April 15, 2023. Solar prominences, which are large-scale structures of plasma suspended in the solar corona, play a critical role in the dynamics of the solar atmosphere and are often associated with coronal mass ejections (CMEs) that can significantly impact space weather. Understanding the physical conditions within prominences, such as temperature, column mass, pressure, and velocity, is essential for advancing our knowledge of solar physics and improving space weather forecasting. 
 
 The Lyman series of hydrogen plays a crucial role in understanding the physical processes occurring in the solar atmosphere. These spectral lines, formed primarily in the chromosphere and transition region, provide valuable diagnostics for plasma properties such as temperature, density, and dynamics. Early studies, such as those of \citet{vernazza1981structure}, established the foundational models of the solar chromosphere and transition region, highlighting the importance of Lyman series emissions in radiative transfer calculations. Subsequent observations from space-based instruments, including SOHO/SUMER \citep{curdt2001sumer}, have significantly advanced our understanding of the spatial and temporal variations of these lines across the solar disk. Recent studies have further explored the diagnostic potential of the Lyman series in solar prominences and fine structures. For instance, \citet{vial2007ly} observed the Lyman $\alpha$ and Lyman $\beta$ profiles in solar prominences using SOHO/SUMER, revealing that the Lyman $\alpha$/Lyman $\beta$ ratio is sensitive to the fine structure and mass motions within prominences. Furthermore, \citet{heinzel2001soho} analyzed the complete hydrogen Lyman spectrum in solar prominences, demonstrating the importance of partial frequency redistribution (PRD) and the role of the prominence-corona transition region (PCTR) in shaping the observed line profiles. In addition, recent work by \citet{hasegawa2020formation} investigated the formation of Lyman $\beta$ and O I 102.7/102.8~nm lines, emphasizing their sensitivity to temperature and velocity perturbations from the low mid-chromosphere to the transition region. These studies highlight the importance of the Lyman series in understanding physical processes in the solar atmosphere.

Recent advancements in prominence modeling have shown that the asymmetries observed in Lyman line profiles, such as Lyman $\alpha$ and Lyman $\beta$, can be attributed to the combined effects of Doppler shifts and optical depth variations within individual fine-structure threads \citep{gunar2008lyman}. These asymmetries are particularly sensitive to the physical parameters of prominences, such as temperature, density, and velocity fields, which vary significantly across the PCTR. The formation of Lyman lines in prominences is strongly influenced by the temperature and density gradients within the PCTR. Studies have shown that the Lyman lines, particularly Lyman $\beta$ and higher members of the series, are formed in the outer layers of the prominence, where the temperature increases sharply from the cool central regions to the hot corona. The brightness temperature of the Lyman lines, derived from the spectral radiance at line center, has been found to be largely independent of the upper level number, indicating that the emitting regions are significantly hotter than the prominence core \citep{stellmacher2008origin}. Furthermore, \citet{tian2009hydrogen} demonstrated that the Lyman line profiles are also influenced by the dynamics of the solar atmosphere, with the Lyman $\alpha$ and Lyman $\beta$ lines exhibiting opposite asymmetries in response to transition-region flows. These highlight the critical role of parameter analysis in interpreting Lyman line observations and diagnosing the physical conditions within prominences.

The primary objective of this study is to analyze the hydrogen Lyman $\beta$ and Lyman $\gamma$ lines of a solar prominence during the pre-eruptive phase and eruptive phase observed by SPICE on April 15, 2023. In particular, their line widths provide diagnostics of the thermal and non-thermal broadening of the line profiles, while their integrated intensities help investigate the spatial variation of the prominence emission.  Understanding these pre-eruptive conditions is crucial, as they set the stage for the subsequent eruption dynamics. This will provide valuable insights into the thermodynamic and dynamic properties of eruptive prominences, contributing to a deeper understanding of their role in solar activity and space weather. The event was observed during its pre-eruptive  and eruptive phases, when significant changes in temperature, density, and pressure are expected, and thus the Lyman line profiles offer unique insight into the evolving prominence plasma.

The structure of this paper is as follows. In Section~\ref{sec:ins}, we describe the SPICE instrument and data set used in this study. In Section~\ref{sec:obs} and Section~\ref{sec:intensity}, we present the prominence observations of April 15, 2023 and extract the Lyman $\beta$ and Lyman $\gamma$ integrated intensity from different spatial regions. In Section~\ref{sec:line_width}, we calculate and analyze the line width of these lines to investigate the thermal and non-thermal broadening of the plasma. In Section~\ref{profile}, we present the spatial and temporal evolution of Lyman $\beta$ and Lyman $\gamma$ line profiles. In Section~\ref{Sec: radial velocity}, we present a new method of radial velocity estimation and discuss the possible Doppler shift.  Finally, in Section~\ref{sec:conclusion}, we summarize and discuss our results.

% The EUI instrument suite on Solar Orbiter is composed of three telescopes, one Full Sun Imager (FSI) and two high-resolution imagers (HRI). The FSI has 2 bands at 17.4~nm and 30.4~nm, centered on line produced by ions  Fe IX-X and He II respectively. The 30.4~nm band is chromospheric and the 17.4~nm band contains also transition region lines. The HRI bands, one at Lyman $\alpha$ and one at 17.4~nm \citep{rochus2020solar}, are comparable to the bands of FSI. With observation of different wavelengths, this instrument enables the acquisition of detailed imaging data of the Sun's entire surface, thereby facilitating scientific analysis of various solar layers and phenomena. Solar Orbiter ran the SOOP (Solar Orbiter Observing Plan) called \text{R\_BOTH\_HRES\_HCAD\_Filaments} on April 15 2023 . 

%\subsection{Imaging channels}
%In this paper, we mainly study prominences by Lyman lines, which are the most prominent lines observed in solar prominences \citep{vial2007ly}. %We analyze Lyman $\beta$ line as a first test.

% \subsection{GONG H$\alpha$ Telescope}\label{sec:h_alpha}

% The Global Oscillation Network Group (GONG) is a network of ground-based telescopes designed to observe the Sun in the H$\alpha$ wavelength (656.3 nm) \citep{harvey1996global}. We use H$\alpha$ images from the El Teide telescope in the GONG network. The H$\alpha$ line is particularly useful for studying solar prominences, filaments, and other chromospheric phenomena, as it provides high-contrast images of these features. 

\section{Observations}

From Global Oscillation Network Group's (GONG) point of view, the prominence was a large quiescent prominence in the northern hemisphere that had been apparent on the disk for 6 days. From 07:30 -10:00 UT the prominence was starting to rise, but was not yet erupting. At around 10:00 UT the central section of the prominence erupted, leaving behind segments to either side.

\subsection{Instrumentation}\label{sec:ins}

\subsubsection{Solar Orbiter}\label{sec:Solar Orbiter}
\begin{figure}
    \centering
    \includegraphics[width=0.50\textwidth]{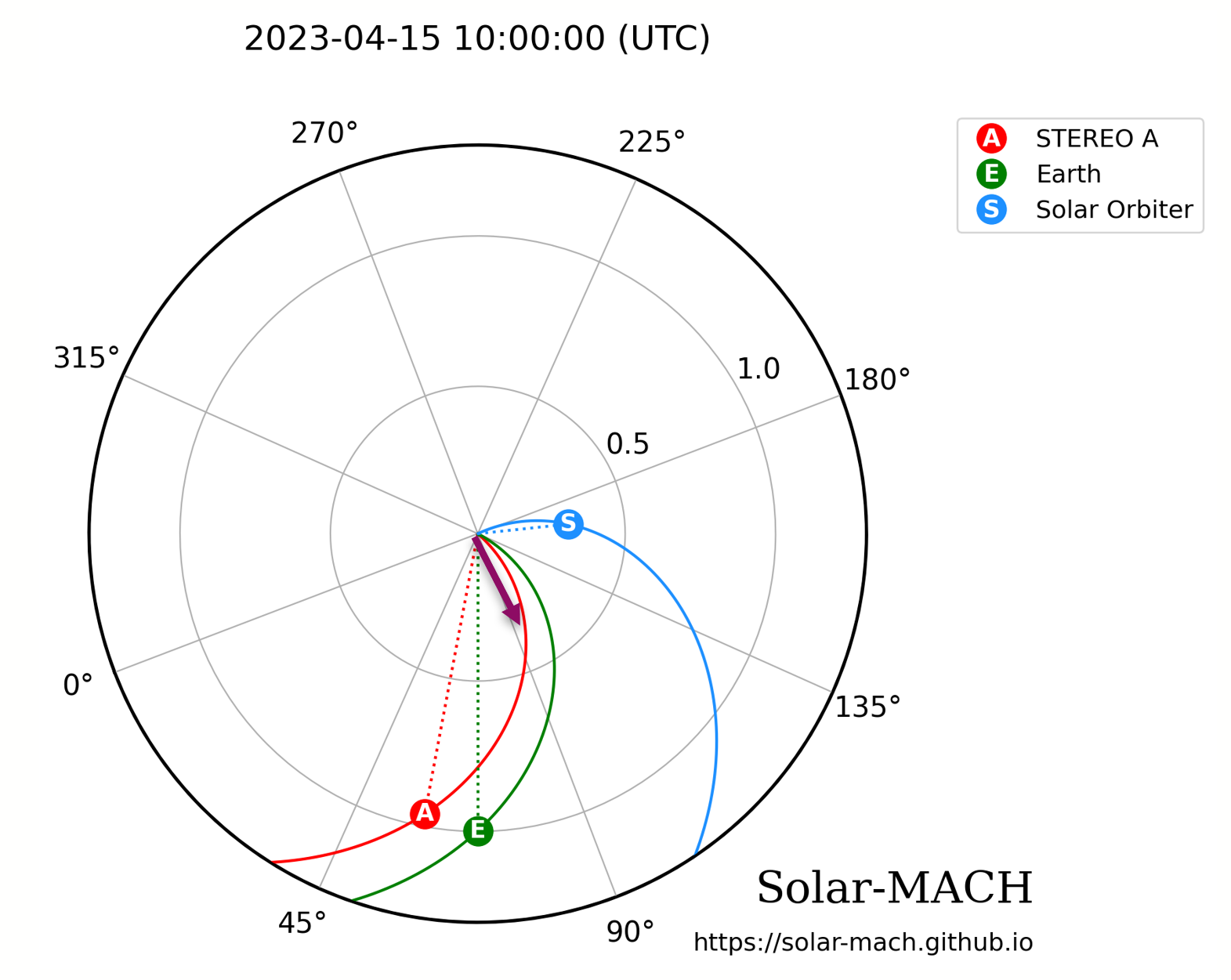}
    \caption{The positions of different spacecraft. The arrow is the eruption direction of the prominence.}
    \label{fig:position}
\end{figure}

Solar Orbiter is a Sun-observing satellite developed by ESA (the European Space Agency)  \citep{muller2020solar}. 
Solar Orbiter operates in a highly elliptical orbit around the Sun, with most of its remote sensing observations taking place at perihelion. As shown in  Figure~\ref{fig:position}, during these observations Solar Orbiter was 0.31~AU from the Sun and roughly in quadrature with Earth. At this time, 1~\arcsec at Solar Orbiter corresponds to 0.31~\arcsec as seen from 1~AU.

The Spectral Imaging of the Coronal Environment (SPICE) instrument is an imaging spectrometer operating at extreme ultraviolet (EUV) wavelengths from 69.6–79.6~nm (short wavelength, SW channel) and 96.0–105.8~nm (long wavelength, LW channel) \citep{2020A&A...642A..14S}. SPICE's primary objective is to support the scientific aims of the Solar Orbiter mission by offering detailed insights into the physical properties and composition of the plasma within the solar atmosphere. In these observations, SPICE contributes to addressing Solar Orbiter's Science Objective on coronal mass ejection (CME) formation through coordinated prominence observations during a dedicated 10-hour perihelion campaign. 
\added{For these observations, SPICE was pointing towards the edge of the solar disk to observe limb prominences and coronal structures above the surface,} with the Earth quadrature configuration, combining SPICE's spectral diagnostics with EUI's high-resolution imaging. We only have EUI/HRI observation till 05:52 UT. In this paper, we concentrate on the analysis of SPICE data. SPICE employs two detector arrays, each with a format of 1024 × 1024 pixels, to simultaneously record dispersed spectra in these bands. The observations utilize a 4\arcsec slit width. The in-flight performances of the instrument are a spatial resolution of 6.7~\arcsec (sampling of 1.098 arcsec/pixel across the slit) and a spectral resolution of about 9.4 pixels for the LW channel \citep{fludra2021first}. The difference in the wavelength sampling between the pre-eruption and eruption datasets (0.0096~nm/pixel and 0.0192~nm/pixel, respectively) results from different binning factors, indicating that the eruption raster was acquired with coarser spectral sampling.

For the pre-eruption raster we analyze, we have a full spectrum in the given range of the LW channel. During the eruptive phase, the raster covers a wider field of view and multiple spectral windows including the Lyman $\beta$ line at 102.57~nm. The data span from 07:03 to 08:09 UT, before the eruption, and from 08:13 to 11:58 UT during the eruption (DOI:10.48326/idoc.medoc.spice.5.0). Table~\ref{tab:obs_params} summarizes the key observational parameters discussed above.

\begin{table*}
\centering
\caption{Summary of SPICE observational parameters in this study (pre-eruption vs. eruption). X and Y coordinates are from the point of view of Solar Orbiter.}
\begin{tabular}{lll}
\hline
Parameter & Pre-eruption & Eruption \\
\hline
\textbf{Observations} & & \\
Date & 15 April 2023 & 15 April 2023 \\
Time window (UTC) & 07:03–08:09 UT & 08:13–11:58 UT \\
SOOP campaign & R\_BOTH\_HRES\_HCAD\_Nanoflares & R\_BOTH\_HRES\_HCAD\_Nanoflares \\
Slit width & 4\arcsec & 4\arcsec \\
Exposure time & 89.6 s & 60.0 s \\

Field of view (Helioprojective) & 
\makecell[l]{X: $-2823\arcsec$ to $-2560\arcsec$ \\ Y: $1035\arcsec$ to $2166\arcsec$} &
\makecell[l]{X: $-3040\arcsec$ to $-2033\arcsec$ \\ Y: $1102\arcsec$ to $2129\arcsec$} \\

\textbf{Spectral characteristics} & & \\
Wavelength range (LW) & 96.0--105.8~nm & 
\makecell[l]{8 selected spectral windows \\ including 102.28--102.87~nm (Ly$\beta$ window)} \\
Spectral resolution & Approximately 0.07~nm & Approximately 0.07~nm \\
Spectral dispersion & 
\makecell[l]{0.0096~nm/pixel (unbinned)} & 
\makecell[l]{0.0192~nm/pixel (binned)} \\
\textbf{Photometric calibration} & & \\
Calibration factor & 100.50~DN/(W m$^{-2}$ sr$^{-1}$ nm$^{-1}$) & 139.65~DN/(W m$^{-2}$ sr$^{-1}$ nm$^{-1}$) \\
\hline
\end{tabular}
\label{tab:obs_params}
\end{table*}

\begin{figure}
    \centering
    \includegraphics[width=0.5\textwidth]{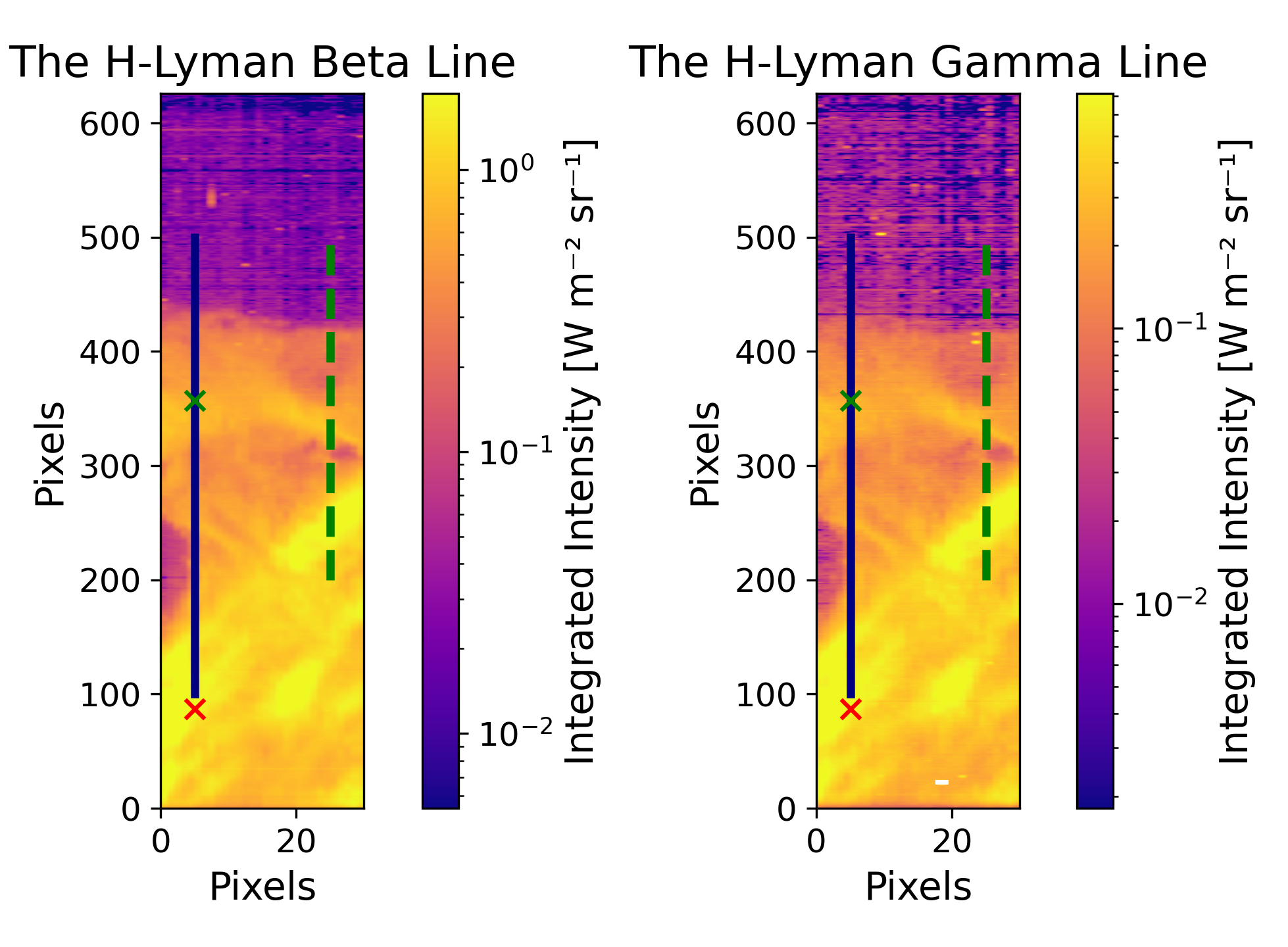}
    \caption{The SPICE observation of the integrated intensity of the Lyman $\beta$ and Lyman $\gamma$ lines taken from  07:03 - 08:09 UT, 15 April 2023. The integrated intensities along the blue cut and the green dashed cut are shown in Figure~\ref{fig:SPICE_slit}. The green cross has the maximum intensity of the prominence region along the blue cut. The red cross has the maximum intensity of the disk region along the blue cut. }
    \label{fig:beta_gamma}
\end{figure}

\begin{figure}
    \centering
\includegraphics[width=0.5\textwidth]{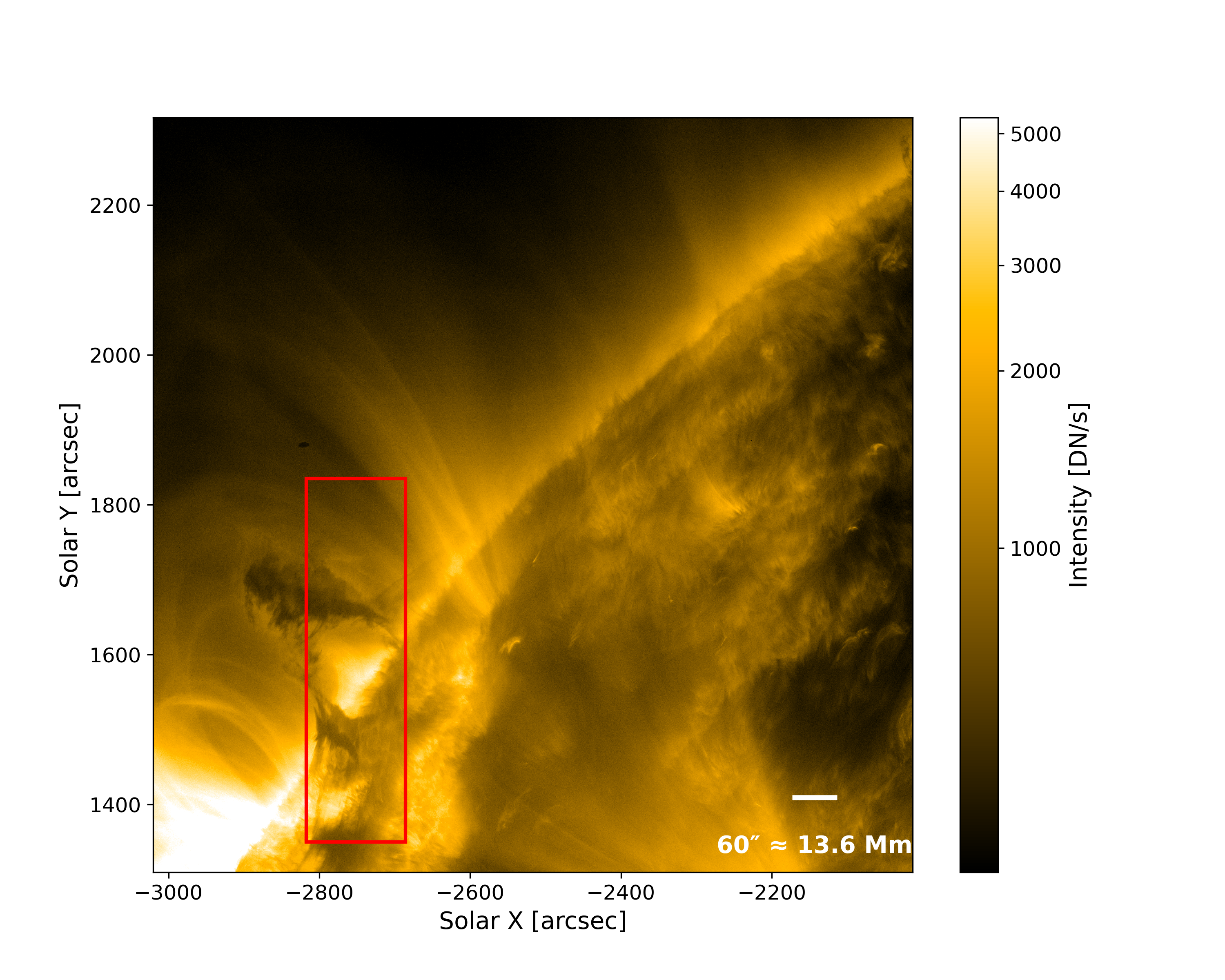}
    \caption{The prominence at 05:52:57 UT, 15 April 2023, observed by EUI/HRI$_\text{EUV}$ 17.4~nm. The red rectangle shows the region observed by SPICE in Figure~\ref{fig:beta_gamma}. The details of structures are slightly different from Figure~\ref{fig:beta_gamma} because of the difference of observation time.}
    \label{fig:euiobs}
\end{figure}

\begin{figure*}
    \centering
    \includegraphics[width=\textwidth]{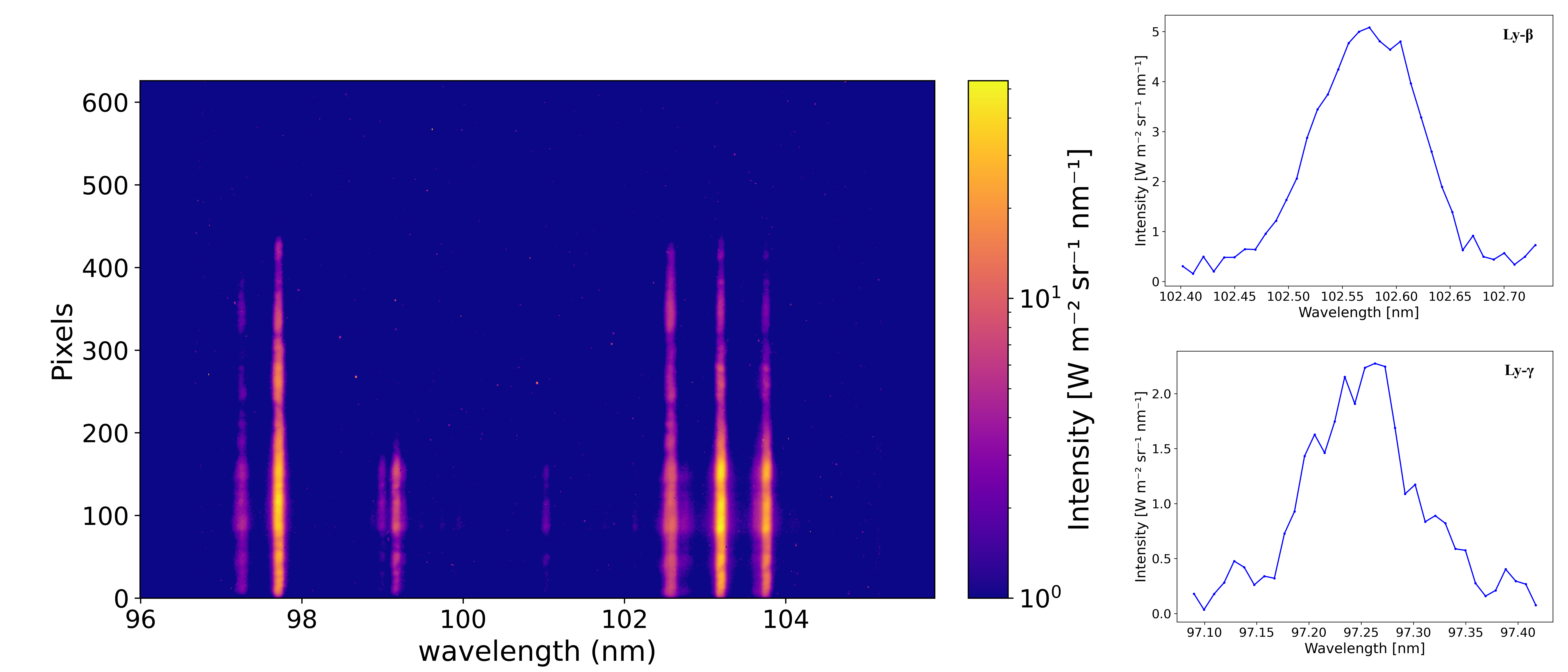}
    \caption{Left: The LW channel observation of SPICE during pre-eruption phase} at 07:37:39 UT. The slit position is the same as the blue cut in Figure~\ref{fig:beta_gamma}. Right: The  Lyman $\beta$ and Lyman $\gamma$ line profiles are from the green cross in Figure~\ref{fig:beta_gamma}.
    \label{fig:lw}
\end{figure*}

\begin{figure*}
    \centering
    \includegraphics[width=\textwidth]{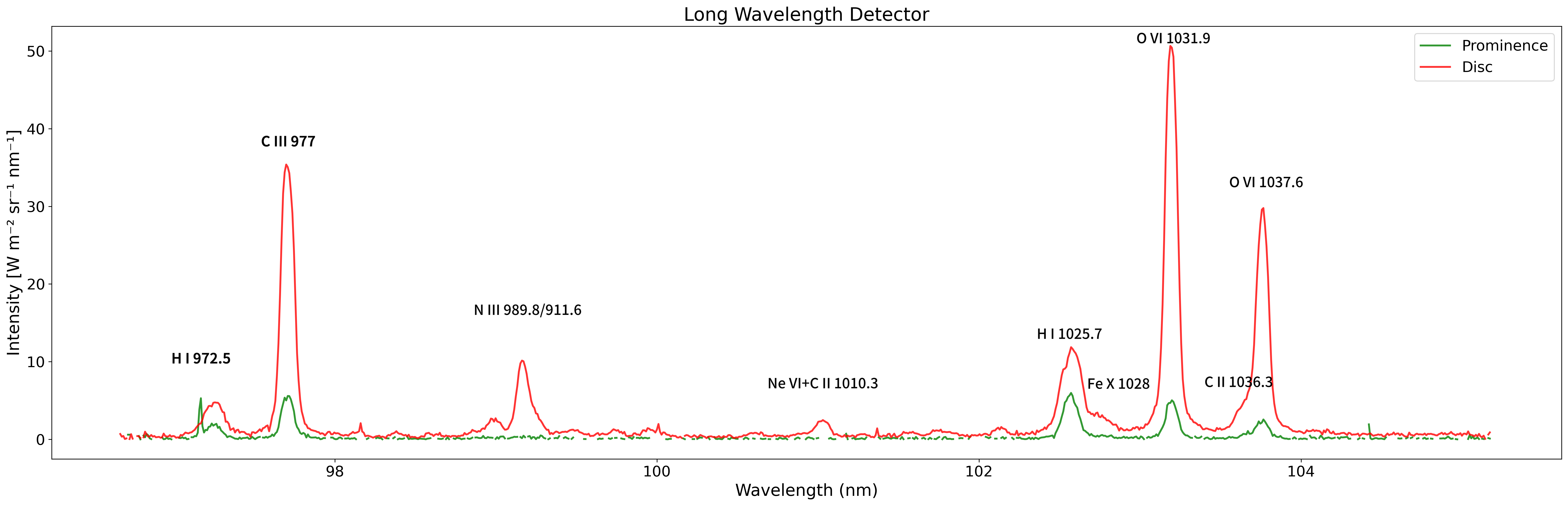}
    \caption{The spectrum of the LW channel during pre-eruption phase at the green cross and the red cross in Figure~\ref{fig:beta_gamma}. Lines wavelengths are in \AA.}
    \label{fig:long_wavelength_spectrum}
\end{figure*}

\begin{figure}
    \centering
    \includegraphics[width=0.40\textwidth]{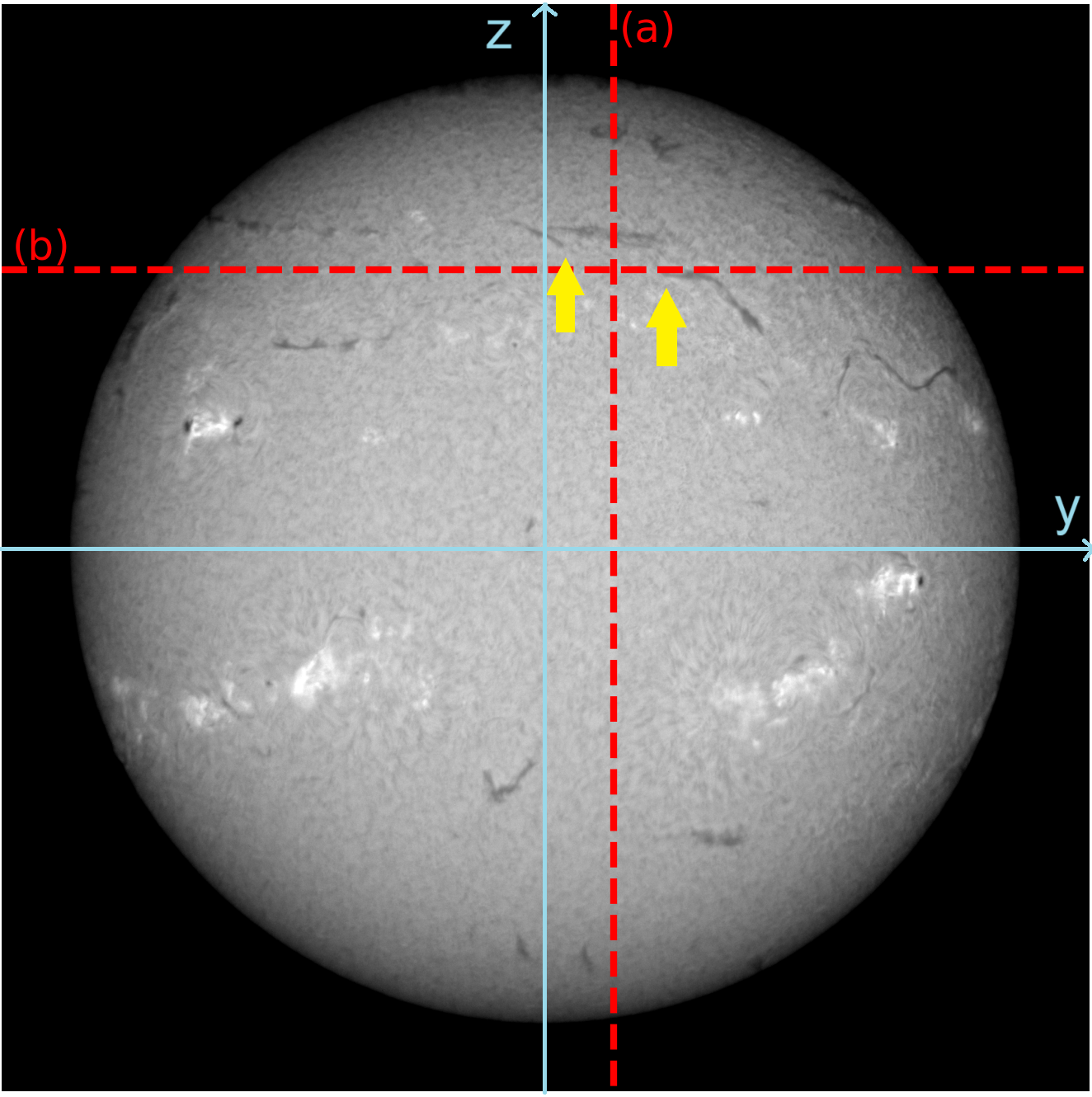}
    \caption{H$\alpha$ image of El Teide at 10:07 UT. The yellow arrows point to the prominence we study. The blue lines show the coordinate we use in the analysis and the red dashed lines show the cross-sections we analyze.}
    \label{fig:H_alpha}
\end{figure}

\begin{figure}
    \centering
    \includegraphics[width=0.5\textwidth]{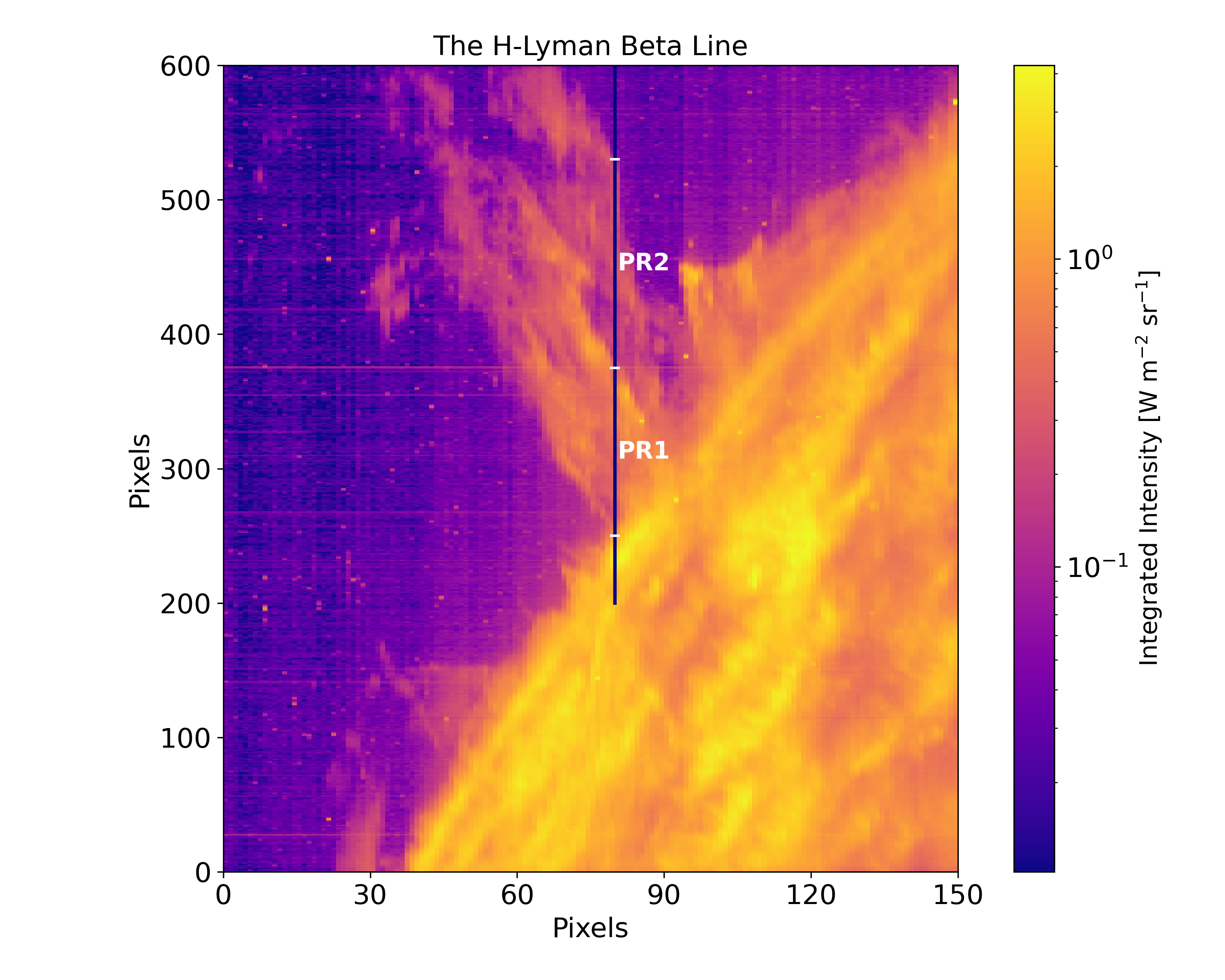}
    \caption{The SPICE observation of the integrated intensity of the Lyman $\beta$ from  08:13 - 11:58 UT, 15 April 2023. 'PR' is the abbreviation of 'Prominence Region'.}
    \label{fig:beta_eruption}
\end{figure}

\subsubsection{GONG H$\alpha$ Telescope}\label{sec:h_alpha}

The Global Oscillation Network Group (GONG) is a network of ground-based telescopes designed to observe the Sun in the H$\alpha$ wavelength (656.3 nm) \citep{harvey1996global}. We use H$\alpha$ images from the El Teide telescope in the GONG network. The H$\alpha$ line is particularly useful for studying solar prominences, filaments, and other chromospheric phenomena, as it provides high-contrast images of these features and has long-term full-disk monitoring. 

\subsubsection{STEREO}
The Extreme UtraViolet Imager aboard the Solar Terrestrial Relations Observatory-A  (STEREO/EUVI-A) \citep{kaiser2008stereo, howard2008sun} produced EUV images in four wavebands. On the day of the observations described here, there was a 2.5 minute cadence in the 19.5~nm band. The response function is peaked at $T = 10^{6.2}~K$ in the 19.5~nm band \citep{wuelser2004euvi}.

\subsection{Observation}\label{sec:obs}

% We have SPICE observation during 01:03-14:58 UT, 15 April 2023, and use pre-eruption rasters from 07:03 UT to 08:09 UT and eruption rasters from 08:13 UT to 11:58 UT} to analyze the event in this paper (DOI:10.48326/idoc.medoc.spice.5.0). 

Figure~\ref{fig:beta_gamma} shows images of integrated intensities of the prominence taken by SPICE before eruption in Lyman $\beta$ and Lyman $\gamma$ lines. Figure~\ref{fig:euiobs} shows the prominence observed by EUI/HRI$_\text{EUV}$ at 05:52:57 UT, 15 April 2023 in the 17.4~nm channel, which is 4 hours before eruption (DOI: https://doi.org/10.24414/z818-4163). The region of the SPICE observation is \added{shown as} the red rectangle region in Figure~\ref{fig:euiobs}. 

Figure~\ref{fig:lw} shows the LW channel observation of SPICE at 07:37:39 UT. The Lyman $\beta$ and Lyman $\gamma$ line profiles on the right panels in Figure~\ref{fig:lw} are from the green cross in Figure~\ref{fig:beta_gamma}. The slit position is the same as the blue cut in Figure~\ref{fig:beta_gamma} (x=5 in pixel coordinate). The green cross is at the location of maximum intensity along the prominence region of the cut. The red cross is at the location of maximum intensity along the disc region of the cut. Figure~\ref{fig:long_wavelength_spectrum} shows the spectrum of the  LW channel at the location of the green and red cross in Figure~\ref{fig:beta_gamma}. In the right side of Figure~\ref{fig:lw}, we can see the  Lyman $\beta$ and Lyman $\gamma$ line profiles from the green cross.

Figure~\ref{fig:H_alpha} shows the H$\alpha$ image of El Teide at 10:07 UT. From GONG H$\alpha$ observations, we know hat the start time of the eruption is around 10:00 UT. 
Figure~\ref{fig:beta_eruption} shows the prominence  in the Lyman $\beta$ line during the eruption. The Lyman $\gamma$ line was not observed during the eruption. From SPICE data we can infer that the altitude range of the prominence during the rasters observing both the pre-eruption and eruption time periods was about $20000~\text{km} \sim 60000~\text{km}$. 
% %
% The slit location is shown with the dashed line in Figure~\ref{fig:beta_gamma}. 

\begin{figure*}
    \centering
    \includegraphics[width=\textwidth]{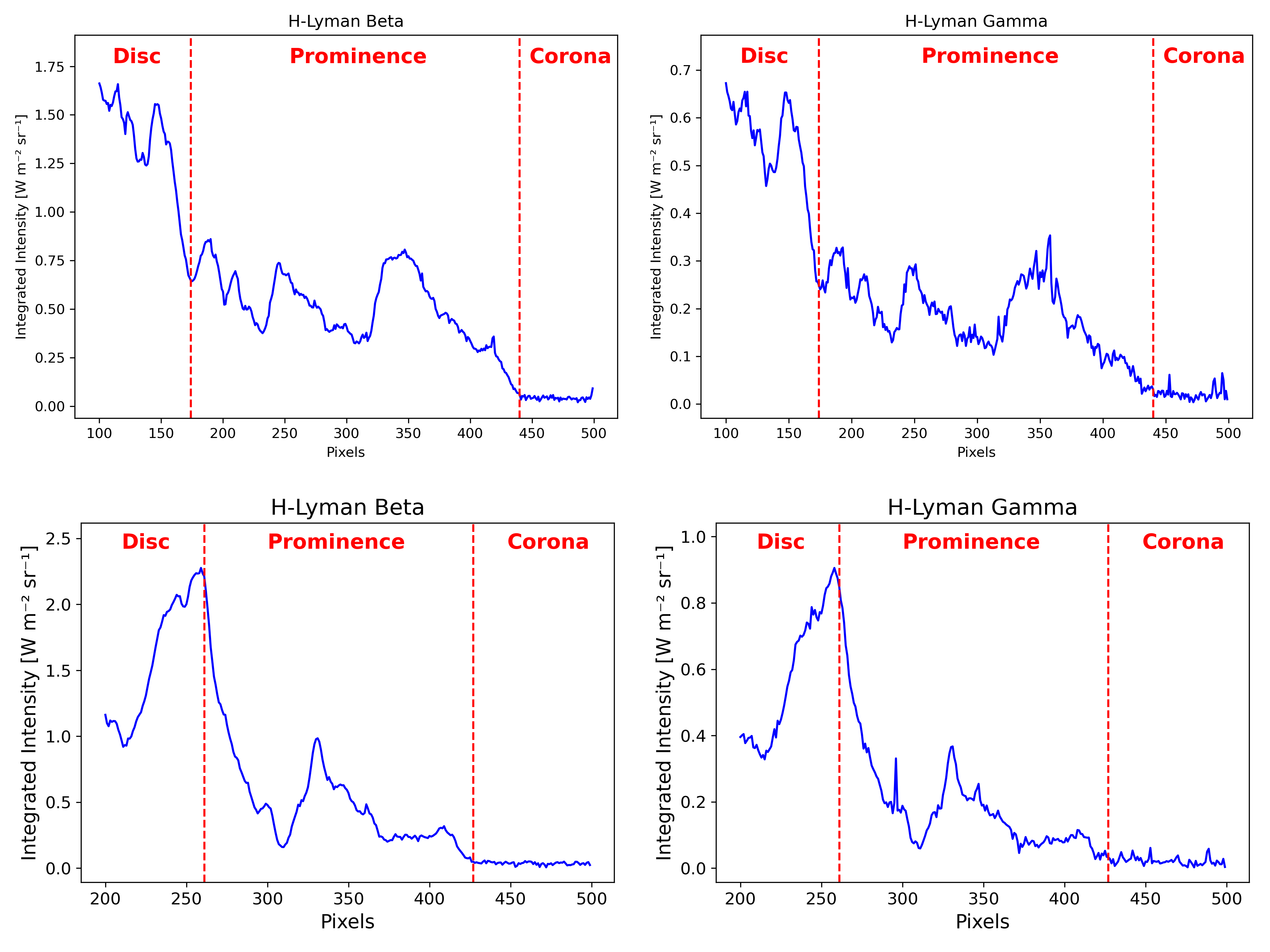}
    \caption{Up: The integrated intensity along the blue cut (x=5) for Lyman $\beta$ and Lyman $\gamma$ lines during pre-eruption phase. Down: The integrated intensity along the green dashed cut (x=25) for Lyman $\beta$ and Lyman $\gamma$ lines during pre-eruption phase.}
    \label{fig:SPICE_slit}
\end{figure*}

\begin{figure*}
    \centering
\includegraphics[width=\textwidth]{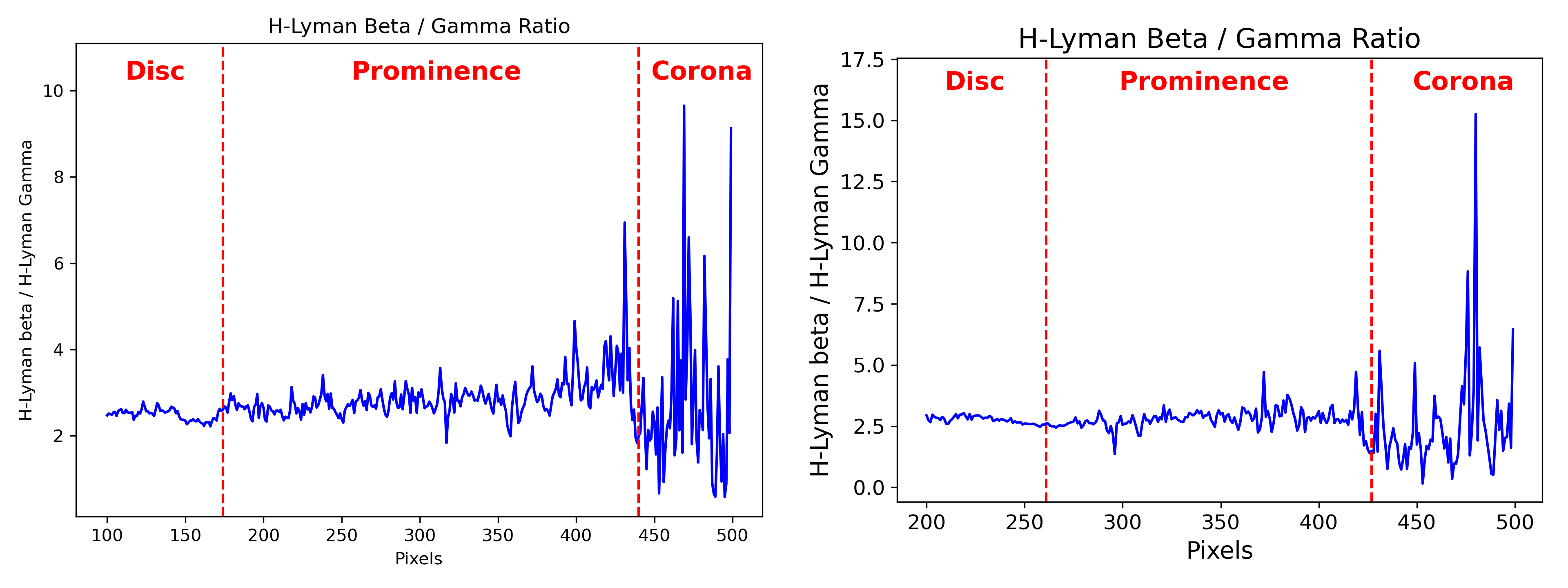}
    \caption{The ratio of the integrated intensity of Lyman $\beta$ and Lyman $\gamma$ lines along the blue cut (x=5, left) and the green dashed cut (x=25, right) during pre-eruption phase.}
    \label{fig:slit_ratio}
\end{figure*}

\begin{figure}
    \centering
    \includegraphics[width=0.5\textwidth]{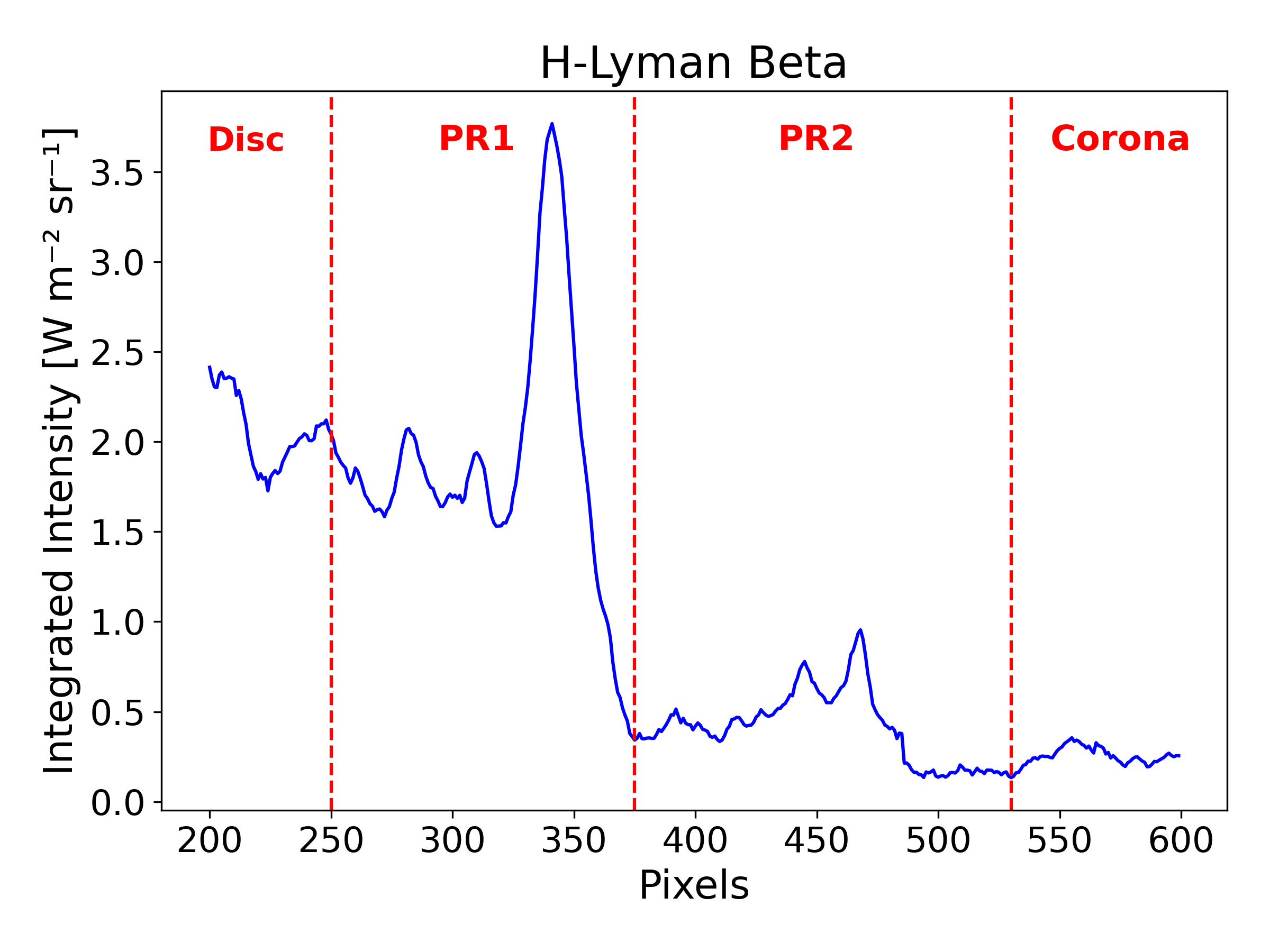}
    \caption{The integrated intensity along the blue cut in Figure~\ref{fig:beta_eruption} for Lyman $\beta$ during eruption phase. PR1 and PR2 are two prominence regions showed in Figure~\ref{fig:beta_eruption}.}
    \label{fig:beta_eruption_slit}
\end{figure}

%

%
% The yellow arrows point to the footpoints of the prominence. We have GONG H$\alpha$ observation from El Teide during 07:30 - 18:34 UT, 15 April 2023. The spacial resolution is 1.063 arcsecond per pixel and the time resolution is 10 min.

\subsection{Integrated Intensity}\label{sec:intensity}

To investigate the spatial variation of the prominence emission, we analyzed the integrated intensity of the Lyman $\beta$ and Lyman $\gamma$ lines along the blue cut indicated in Figure~\ref{fig:beta_gamma}. The  intensity is obtained by \added{performing a standard integration of} the spectral intensity over 0.17~nm from central wavelength of each line at every spatial pixel along the cut. The value of 0.17~nm was chosen as it corresponds to roughly half the width of the Lyman $\beta$ window in data from the eruption raster, ensuring we capture the full profile. We do not use Gaussian fitting because the line profiles of Lyman $\beta$ and Lyman $\gamma$ deviate from Gaussian distributions due to the joint influence of noisy data, their intrinsically non-Gaussian shape when the lines are optically thick, and blending with other lines. The He~II 102.527~nm line has a contribution about 1.5~\% to the integrated intensity of Lyman $\beta$ line \citep{ebadi2009he}. In this observation, actually, we cannot see the He~II 102.527~nm line in most Lyman $\beta$ line profiles. 

The integrated intensity along the blue cut in Figure~\ref{fig:beta_gamma} for Lyman $\beta$ and Lyman $\gamma$ lines is shown in the upper panel of Figure~\ref{fig:SPICE_slit}. The solar disc region is on the left side of the first red dashed line and it shows relatively high intensity. The solar corona region is on the right side of the second red dashed line and it shows a very low intensity. This mainly reflects the higher temperature in the corona region that ionizes hydrogen or much higher density of the plasma in the disc region compared with the corona region. In the lower panel of Figure~\ref{fig:SPICE_slit}, we show the integrated intensity along the green dashed cut in Figure~\ref{fig:beta_gamma}. From Figure~\ref{fig:euiobs} and Figure~\ref{fig:beta_eruption}, we can infer that the angle of the prominence \added{axis }with \added{the} LOS of Solar Orbiter is small and the position of the farther away footpoint can be quite far away from the closer footpoint. Prominence materials along the green cut are behind most prominence materials along the blue cut. The projection effects along the line of sight should be different for the blue cut and the green dashed cut.

\begin{table*}
\centering
\caption{Mean integrated intensities and mean line widths of the Lyman $\beta$ and Lyman $\gamma$ lines measured in the disk and prominence regions before and during the eruption.}
\label{tab:lyman_intensity_width}
\begin{tabular}{llcccc}
\hline
Phase & Region & Line & Integrated Intensity & Observed Line Width & Intrinsic Line Width \\
      &        &      & (W m$^{-2}$ sr$^{-1}$) & (nm) & (nm) \\
\hline
Pre-eruption & Disk       & Lyman $\beta$  & 1.29 & 0.12 & 0.09 \\
Pre-eruption & Disk       & Lyman $\gamma$ & 0.47 & 0.15 & 0.13 \\
Pre-eruption & Prominence & Lyman $\beta$  & 0.52 & 0.13 & 0.10 \\
Pre-eruption & Prominence & Lyman $\gamma$ & 0.19 & 0.18 & 0.16 \\
Eruption     & Disk       & Lyman $\beta$  & 1.50 & 0.09 & 0.05 \\
Eruption     & Prominence & Lyman $\beta$  & 1.07 & 0.09 & 0.05 \\
\hline
\end{tabular}
\label{table:summary}
\end{table*}

In Figure~\ref{fig:slit_ratio}, we can see that there is a roughly constant ratio for the integrated intensity of Lyman $\beta$ and Lyman $\gamma$ lines in the Disc and the Prominence regions, which is around 2.5. This relatively stable ratio suggests that both lines are formed under similar radiative transfer conditions in these regions. It is a possible indication that both lines are formed through resonance scattering of the chromospheric radiation. In \cite{heinzel2001soho}, the ratio for the integrated intensity of Lyman $\beta$ and Lyman $\gamma$ lines in the Prominence regions is from 2.22 to 3.03. Our result is within this range. Also, the ratio of peak values of Lyman $\beta$ and Lyman $\gamma$ lines profiles can be derived from the SUMER atlas of \cite{parenti2004prominence,parenti2005prominence} , which provides a value around 5 for the quiet sun and the prominence region. In \cite{ebadi2009he} , the ratio of the peak values of Lyman $\beta$ and Lyman $\gamma$ lines profiles is around 1.6. Since the shapes of Lyman $\beta$ and Lyman $\gamma$ lines profiles are almost identical for each observation, we can assume the ratio of peak intensities is the same as ratios of integrated intensities for Lyman $\beta$ and Lyman $\gamma$ lines. So we can compare results in \cite{parenti2004prominence,parenti2005prominence} and \cite{ebadi2009he} with \cite{heinzel2001soho} and our results. Results in \cite{parenti2004prominence,parenti2005prominence,ebadi2009he} are different from results from \cite{heinzel2001soho} and ours. From \cite{lemaire2012solar} , we know that the Lyman $\alpha$ to Lyman $\beta$ irradiance ratio changes with solar activity. During minimum activity period, the standard deviation is around 30~\%; during moderate activity period, the standard deviation is around 50~\%. There is still no statistic for the solar maximum period. These differences of ratios among \cite{parenti2004prominence,parenti2005prominence} , \cite{ebadi2009he} , \cite{heinzel2001soho} and us can be explained if the Lyman $\beta$ to Lyman $\gamma$ peak intensity ratio follows similar variations as the Lyman $\alpha$ to Lyman $\beta$ irradiance ratio. Our observations are from 2023. Observations of \cite{heinzel2001soho} and \cite{parenti2004prominence,parenti2005prominence} are from 1999. These are during solar moderate activity period. The values should be within the normal variation range ($\sim$50~\%). The value from \cite{ebadi2009he} is quite low compared with other papers. This is because their Lyman $\beta$ line observation is from 2004 and their Lyman $\gamma$ line observation is from 2001. The year of 2001 is during solar maximum. Both Lyman $\beta$ and Lyman $\gamma$ lines intensities in 2004 should be much lower than that in 2001 because of the reduction of the solar activity. Therefore, if we use Lyman $\beta$ line observation in 2004 and Lyman $\gamma$ line observation in 2001 to calculate the ratio, the value will be much lower than values calculated within the same year.

The ratio varies a lot in the Corona region in Figure~\ref{fig:slit_ratio}. Such a large variation is likely due to the extremely low signal level in the corona, where noise, scattered light, small changes in optical thickness, or weak emission can strongly affect the ratio.

We also investigate the spatial variation of the prominence emission of the Lyman $\beta$ line during eruption phase. In Figure~\ref{fig:beta_eruption_slit}, we can see how integrated intensity changes along the blue cut in Figure~\ref{fig:beta_eruption}. In Figure~\ref{fig:beta_eruption}, some pixels in prominence even have larger intensities than the disc region. This suggests that during the eruption, additional heating or increased column density of hydrogen can temporarily make the prominence brighter than the solar disc in Lyman $\beta$ line. From Figure~\ref{fig:beta_eruption}, we can see that Prominence Region 1 and Prominence Region 2 seem to belong to different parts located at different positions with respect to the plane of sky. The integrated intensity of prominence region 1 is larger than prominence region 2. The difference may come from projection effects and different optical thickness along the line of sight. The integrated intensity of the Lyman $\beta$ line in prominence region 1 during the eruption phase is generally larger than that during the pre-eruption phase. This is consistent with the expectation that eruption processes lead to enhanced plasma heating, mass motions, and possibly compression, all of which can increase line emissivity. But the integrated intensity of the Lyman $\beta$ line in prominence region 2 is even lower than that during the pre-eruption phase. One plausible explanation is that the eruptive prominence region 2 is a low density region. 

% \added{In general, the integrated intensity of the eruptive prominence is much higher than the integrated intensity of the quiescent prominence shown in} \cite{heinzel2001soho}\added{. In} \cite{heinzel2001soho}\added{, the mean integrated intensities of several prominences are 0.22 - 0.70 W m$^{-2}$ sr$^{-1}$ for Lyman $\beta$ line and 0.04 - 0.18 W m$^{-2}$ sr$^{-1}$ for Lyman $\gamma$ line. The mean integrated intensities of the prominence region of two slits we study during pre-eruption phase are 0.52 W m$^{-2}$ sr$^{-1}$ for Lyman $\beta$ line and 0.19 W m$^{-2}$ sr$^{-1}$ for Lyman $\gamma$ line. The values are close to the May 28, 1999 prominence mentioned in} \cite{heinzel2001soho}\added{, which is low and dense. During the eruption phase, the mean integrated intensity in Lyman $\beta$ of the prominence region of the blue cut we study is 1.07 W m$^{-2}$ sr$^{-1}$. This value is obviously larger than all the quiescent prominences shown in \cite{heinzel2001soho}. Enhanced plasma heating, mass motions, and possibly compression during eruption can be reasons for the increased integrated intensity.}

\begin{figure}
    \centering
\includegraphics[width=0.5\textwidth]{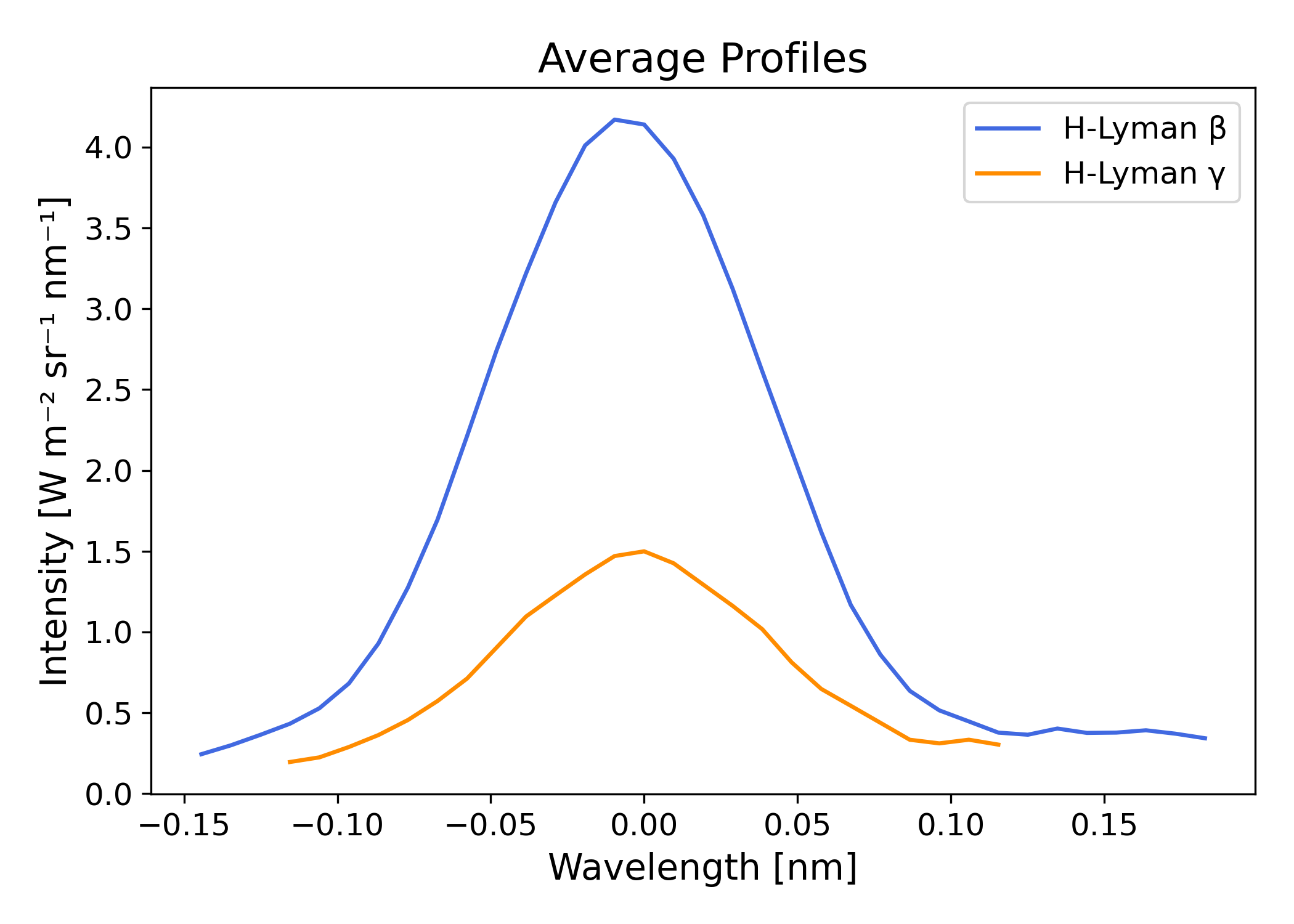}
    \caption{The average profiles of Lyman $\beta$ and Lyman $\gamma$ lines of Prominence Region during pre-eruption phase along the green dashed cut in Figure~\ref{fig:beta_gamma}. The center wavelength is 102.57~nm and 97.25~nm, respectively.}
    \label{fig:average_profile}
\end{figure}

\begin{figure}
    \centering
\includegraphics[width=0.5\textwidth]{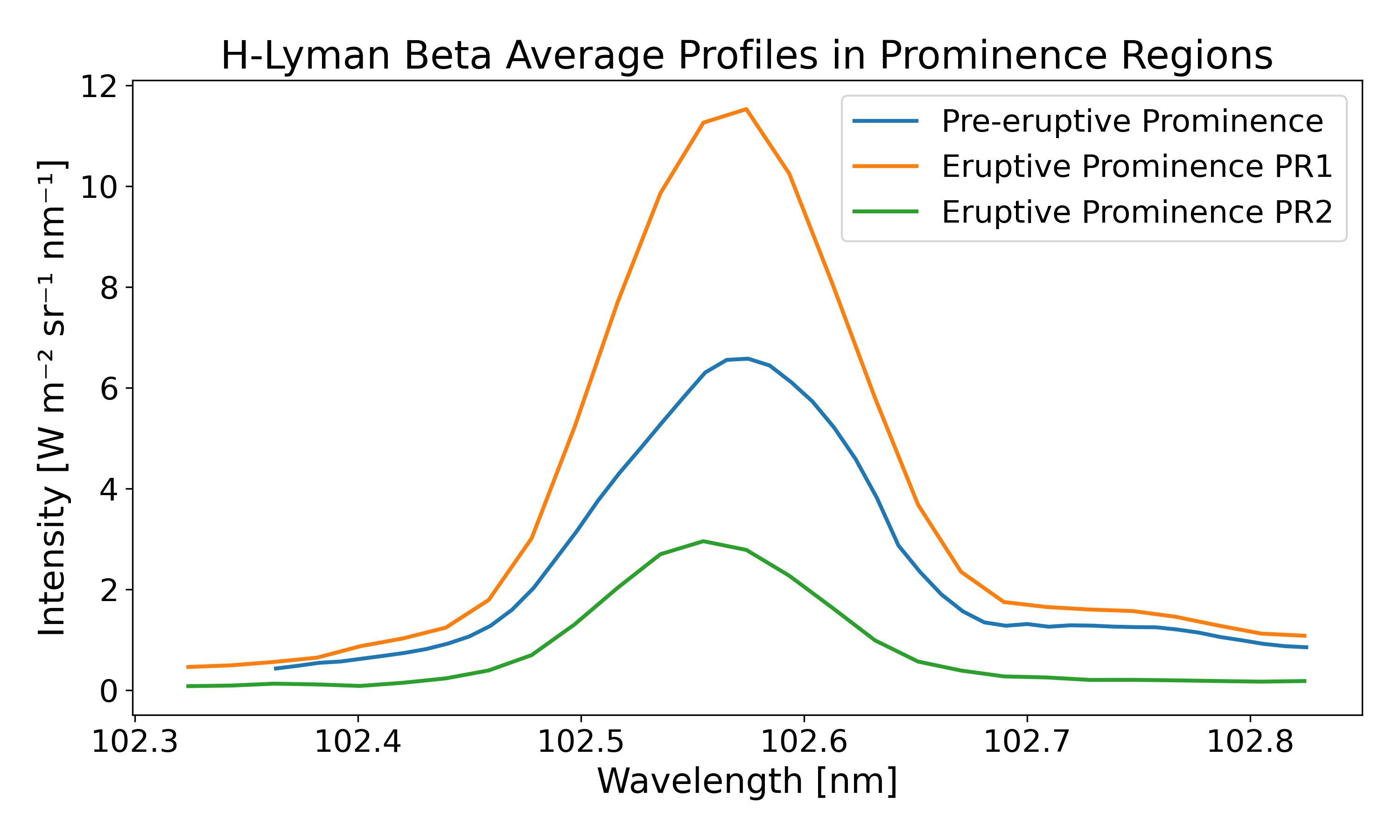}
    \caption{The average profiles of Lyman $\beta$ in the prominence region along the blue cuts during the pre-eruption phase and eruption phase.}
    \label{fig:beta_prominence_profiles}
\end{figure}

If we compare the Lyman $\beta$ line profiles of Figure~\ref{fig:average_profile} and Figure~\ref{fig:beta_prominence_profiles} during the pre-eruptive phase, we can find that the peak values of Lyman $\beta$ line profiles along the blue cut is about 1.5 times larger than the green cut. The peak value of Lyman $\beta$ line is around 6.5~W~m$^{-2}$~sr$^{-1}$~nm$^{-1}$ along the blue cut and 4.2~W~m$^{-2}$~sr$^{-1}$~nm$^{-1}$ along the green cut . As discussed above, the difference may come from projection effects along the line of sight as the two cuts belong to different parts located at different positions with respect to the plane of sky. In \cite{parenti2004prominence,parenti2005prominence}, the peak value of Lyman $\beta$ line in prominence regions is also around 6.5~W~m$^{-2}$~sr$^{-1}$~nm$^{-1}$.

We have provided mean integrated intensities and mean line widths of the Lyman $\beta$ and Lyman $\gamma$ lines measured in the disk and prominence regions before and during the eruption in Table~\ref{table:summary}. In general, the integrated intensity of the eruptive prominence is much higher than the integrated intensity of the quiescent prominence shown in \cite{heinzel2001soho}. In \cite{heinzel2001soho}, the mean integrated intensities of several prominences are 0.22--0.70~W~m$^{-2}$~sr$^{-1}$ for Lyman $\beta$ line and 0.04--0.18~W~m$^{-2}$~sr$^{-1}$ for Lyman $\gamma$ line. The mean integrated intensities of the prominence region of two cuts we study during the pre-eruption phase are 0.52~W~m$^{-2}$~sr$^{-1}$ for Lyman $\beta$ line and 0.19~W~m$^{-2}$~sr$^{-1}$ for Lyman $\gamma$ line. The values are close to the May 28, 1999 prominence mentioned in \cite{heinzel2001soho}, which is low and dense. In \cite{vial2007ly}, the integrated intensities of the prominence region have ranges of 0.027--0.237~W~m$^{-2}$~sr$^{-1}$ for Lyman $\beta$ line. Values of \cite{vial2007ly} come from the SUMER observation in 2005, which is  closer to solar minimum. On the other hand, our observations and those of \cite{heinzel2001soho} are closer to solar maximum. This could explain why the integrated intensities of the prominence region for Lyman $\beta$ line in \cite{vial2007ly} are much smaller. The 8th Orbiting Solar Observatory (OSO-8) observation in \cite{vial1982optically} shows that the integrated intensities of the prominence region for Lyman $\beta$ line are 0.44--0.55~W~m$^{-2}$~sr$^{-1}$. The OSO-8 observation in \cite{vial1982optically} is also during solar minimum, but the values are far from \cite{vial2007ly}. As mentioned in \cite{vial2007ly}, this can be explained by the fact that the structures studied in \cite{vial1982optically} are very different from those in \cite{vial2007ly}. During the eruption phase, the mean integrated intensity in Lyman $\beta$ of the prominence region of the blue cut we study is 1.07~W~m$^{-2}$~sr$^{-1}$. This value is obviously larger than all the quiescent prominences shown in \cite{heinzel2001soho}, \cite{vial2007ly} and \cite{vial1982optically}. Enhanced plasma heating, mass motions, and possibly compression during eruption can be reasons for the increased integrated intensity.

% \begin{figure}
%     \centering
% \includegraphics[width=0.5\textwidth]{beta_average_profile_eruption.png}
%     \caption{The average profiles of Lyman $\beta$ in the two prominence regions during the eruption phase in Figure~\ref{fig:beta_eruption_slit}.}
%     \label{fig:beta_average_profile_eruption}
% \end{figure}

\subsection{Line Width}\label{sec:line_width}

\begin{figure}
    \centering
\includegraphics[width=0.5\textwidth]{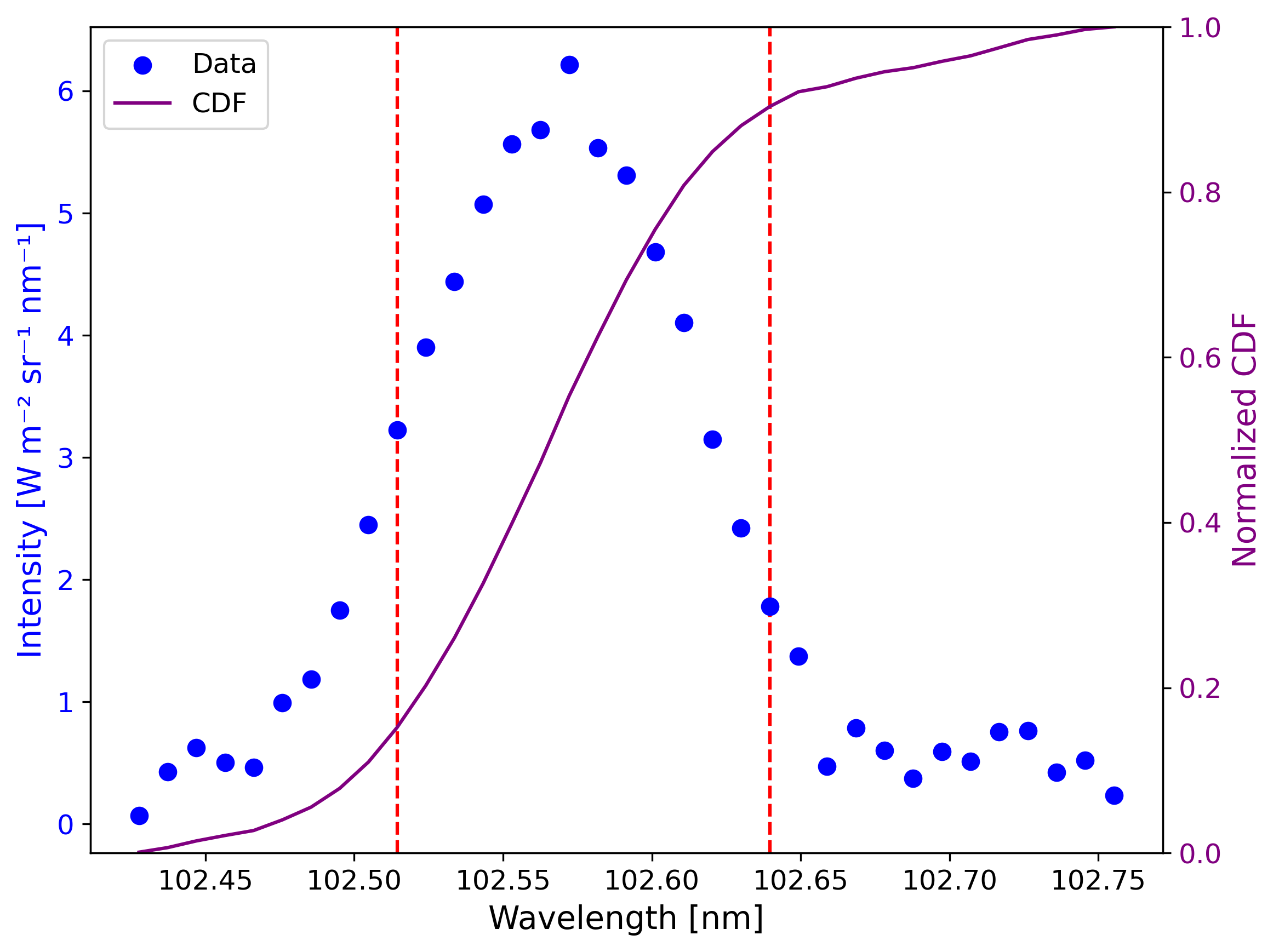}
    \caption{The method of line width calculation. CDF is the cumulative distribution function. The two red dashed lines lie in the locations of CDF=0.12 and CDF=0.88. The distance of the two red dashed lines is the line width of this line profile. The blue dots show data points and CDF is shown by the purple line.}
    \label{fig:line_width_calculation}
\end{figure}

Here we discuss the properties of \added{the observed profiles without performing a full deconvolution} from the instrumental profile because we don't have \added{enough} information about the instrumental profile of the SPICE instrument. \added{We provide values for the measured line widths before and after applying a simple correction for the instrumental width.} The average profiles of Lyman $\beta$ and Lyman $\gamma$ lines in the Prominence Region along the green dashed cut (x=25) in Figure~\ref{fig:beta_gamma} during the pre-eruption phase are shown in Figure~\ref{fig:average_profile}. In general, the line profile of Lyman $\beta$ have higher intensity and wider line shape than the line profile of Lyman $\gamma$. 

In Figure~\ref{fig:beta_prominence_profiles}, we can see that the line profiles of eruptive prominence region 1 have generally larger intensity and wider line shape than that during the pre-eruption phase. This may indicate that eruptive prominence region 1 has either higher density or higher temperature along the line of sight. The broadening could be related to the change of temperature among different rasters or motions along LOS. In Figure~\ref{fig:beta_prominence_profiles}, we also compare the average profiles of Lyman $\beta$ in the prominence region during the pre-eruption phase and eruption phase. The eruptive prominence region 1 has higher intensity and broader line width than pre-eruption phase, which is as we expected.  But the eruptive prominence region 2 has lower intensity and narrower line width than pre-eruption phase. The eruptive prominence region 2 also shows some blue-shift in Lyman $\beta$ line profile.

We calculate the line width of the Lyman $\beta$ and Lyman $\gamma$ profiles from SPICE in order to diagnose the plasma conditions, since the line width carries information on both thermal and non-thermal broadening. To estimate the line widths of the Lyman $\beta$ and Lyman $\gamma$ lines we adopt the quantile method, which has been successfully applied in recent IRIS studies \citep{kerr2015iris,ruan2018dynamic}. This method relies on constructing the cumulative distribution function (CDF) of the line profile. The line center is then given by the 50\% quantile. Compared with standard Gaussian fitting, the quantile method is more robust against noise and line asymmetries, and thus particularly well-suited for our data where the observed line profiles deviate from a simple Gaussian shape.

The line width measurement by the quantile method can be chosen so that it conveniently reproduce the full-width at half-maximum of a Gaussian profile. We solve the equation derived from the ratio of two functions involving a Gaussian distribution with a mean $\mu = 0$ and calculate the corresponding CDF values. The probability density function (PDF) of a Gaussian distribution with mean \(\mu = 0\) and standard deviation \(\sigma\) is given by:

\begin{equation}
f(x) = \frac{1}{\sigma \sqrt{2\pi}} \exp\left(-\frac{x^2}{2\sigma^2}\right)
\end{equation}

$x$ is the variable that follows Gaussian distribution. At half maximum intensity, $f(x)/f(0)=0.5$, so

\begin{equation}
   \exp\left(-\frac{x^2}{2\sigma^2}\right) = 0.5
\end{equation}

This could be simplified as

\begin{equation}
x^2 + 2\sigma^2 \ln(0.5) = 0
\label{line_width_posi}
\end{equation}

The Gaussian CDF for a given \(x\) is defined as:

\begin{equation}
\text{CDF}(x) = \frac{1}{2} \left(1 + \text{erf}\left(\frac{x-\mu}{\sqrt{2} \sigma}\right)\right)
\end{equation}
\label{erf}
   
In Equation~\ref{erf}, erf is error function. After substituting the solutions for $x$ in Equation~\ref{line_width_posi} into this CDF formula, we can find the corresponding CDF values at the half maximum intensity positions. The results are 0.12 and 0.88.

\begin{figure}
    \centering
\includegraphics[width=0.5\textwidth]{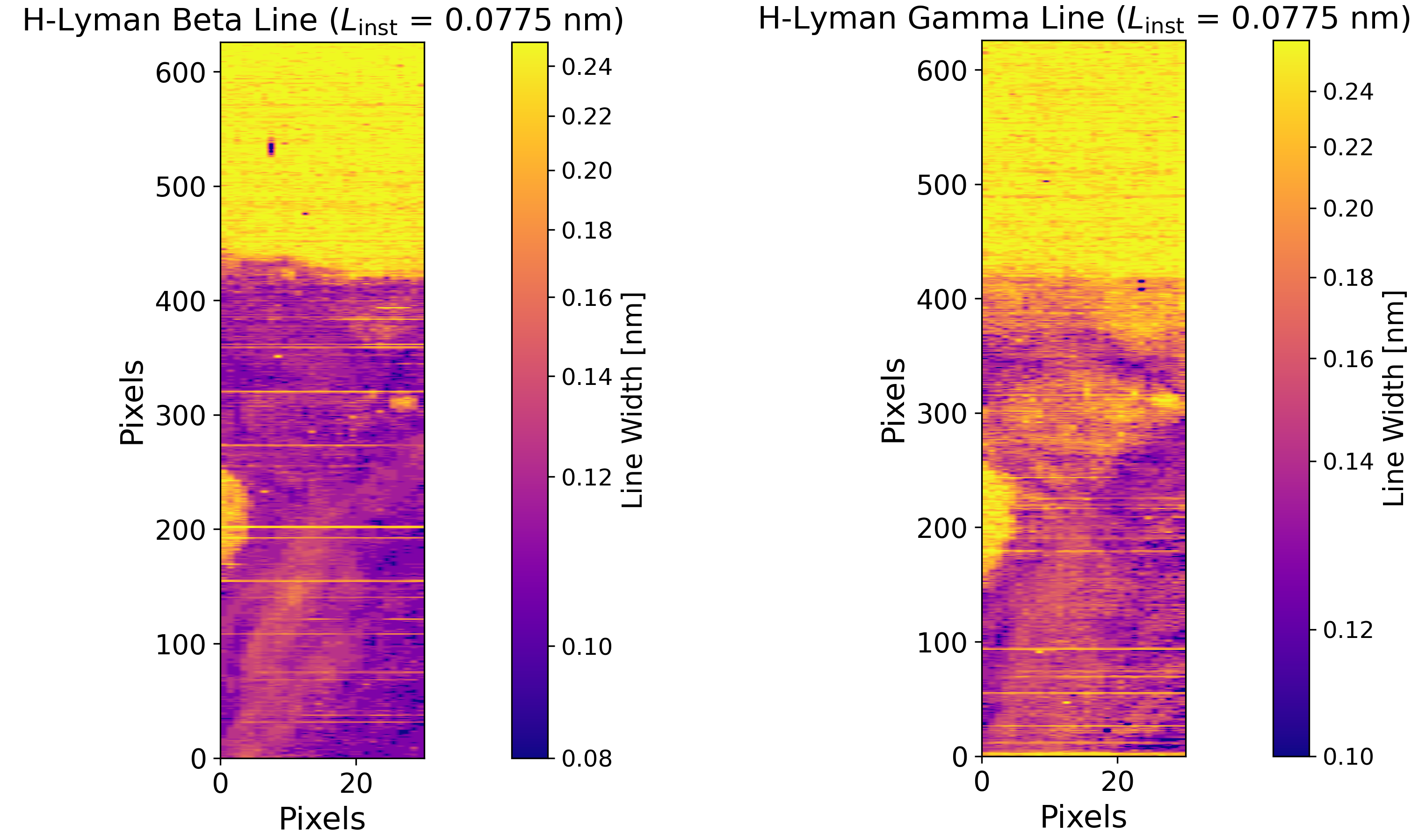}
    \caption{The line width of Lyman $\beta$ and Lyman $\gamma$ lines profiles during pre-eruption phase.}
    \label{fig:SPICE_line_width}
\end{figure}

By identifying the locations of CDF=0.12 and CDF=0.88, we get the line width (Figure~\ref{fig:line_width_calculation}). The line width before the eruption for the Lyman $\beta$ and Lyman $\gamma$ lines is shown in Figure~\ref{fig:SPICE_line_width}. Considering the instrumental broadening $\text{L}_\text{inst}$, values that are lower than $\text{L}_\text{inst}$ have been changed to the value of $\text{L}_\text{inst}$.

The average line widths in the prominence region (y-pixel 326 to 426) are 0.13~nm for the Lyman $\beta$ line and 0.18~nm for the Lyman $\gamma$ line. The average line widths in the disk region (y-pixel 0 to 100) are 0.12~nm for the Lyman $\beta$ line and 0.15~nm for the Lyman $\gamma$ line. We can calculate the intrinsic line width after accounting for the instrumental broadening ($\text{L}_\text{inst}$ = 0.0775 nm) of SPICE's long-wavelength channel \citep{young2025fe}. The average line widths in the prominence region are 0.10~nm for the Lyman $\beta$ line and 0.16~nm for the Lyman $\gamma$ line after correction. The average line widths in the disk region are 0.09~nm for the Lyman $\beta$ line and 0.13~nm for the Lyman $\gamma$ line after correction. The broader width of the Lyman $\gamma$ line compared to Lyman $\beta$ can be attributed to the \added{smaller signal-to-noise ratio of the Lyman $\gamma$ line. The noisy data can affect the calculation of the line width and usually enlarges the value. The effect is larger in the Lyman $\gamma$ line because the intensity of the Lyman $\gamma$ line is smaller.}

% The intrinsic width of the Lyman~$\gamma$ line is found to be slightly larger than that of Lyman~$\beta$ both on the disk and in the prominence. This difference can be explained by several physical factors. 
% The most likely explanation for the larger intrinsic width of Lyman~$\gamma$ compared with Lyman~$\beta$, when both lines originate from the same prominence structure, is differences in radiative transfer and opacity effects rather than formation-height differences. In particular, Lyman~$\beta$ is typically more optically thick and may suffer from core saturation and multiple scattering that modifies the apparent profile (flattening the core and reducing fitted widths), whereas Lyman~$\gamma$—being less thick—samples a broader range of velocities and retains more pronounced wings. Additional contributions from cascade excitation and partial redistribution (PRD) effects can further broaden the Lyman~$\gamma$ profile. Secondary possibilities include Stark (pressure) broadening affecting higher Lyman members and unresolved blends in the Lyman~$\gamma$ wings.

This result shows that the intrinsic line width of the prominence and disk regions are generally similar. This suggests similar line formation conditions, like temperature, and optical thickness through the opacity, for the two different regions. 

\begin{figure}
    \centering
\includegraphics[width=0.5\textwidth]{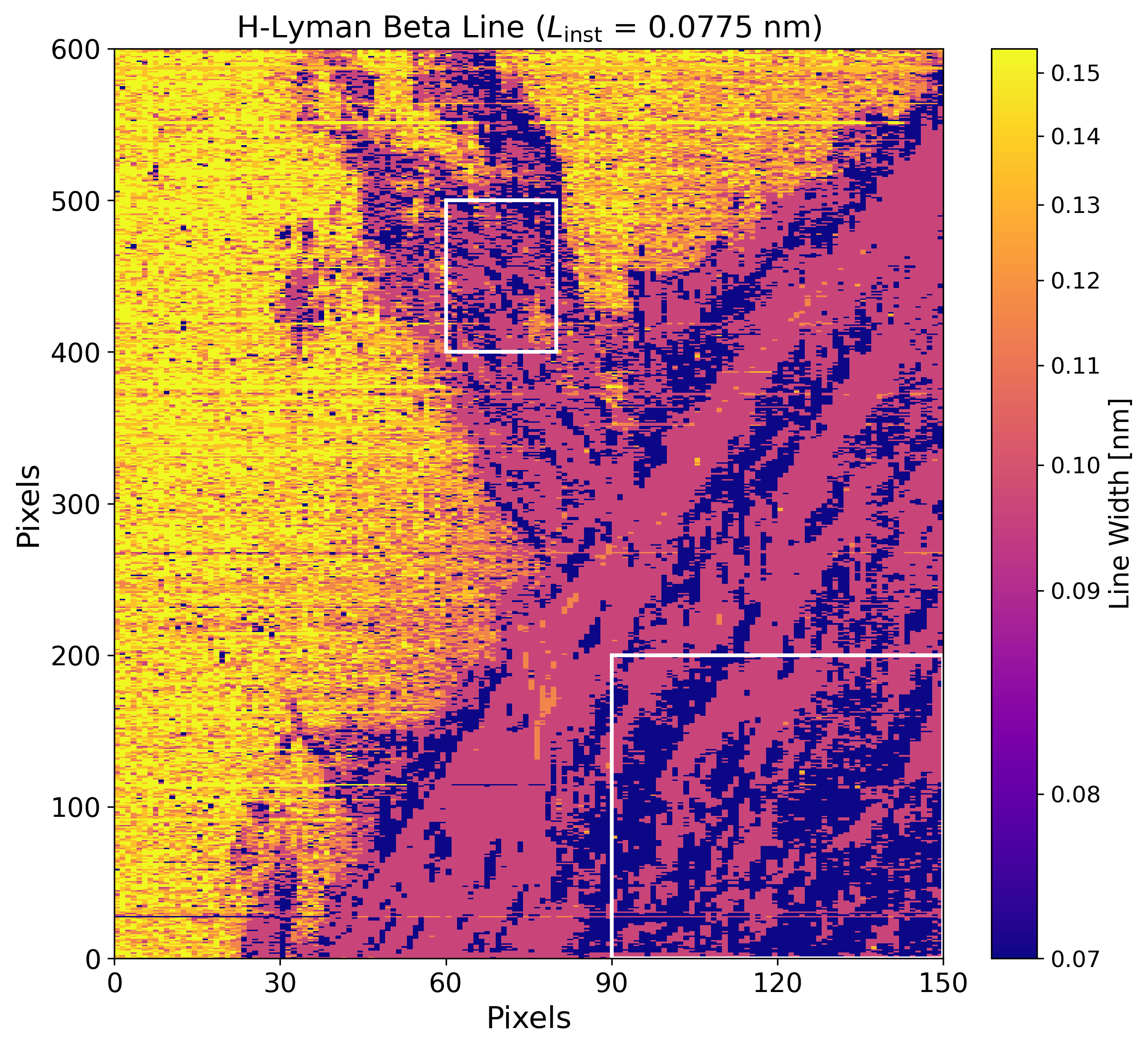}
    \caption{The line width of Lyman $\beta$ lines profiles during eruption phase. The white rectangles show locations where we take samples for the disk region and the prominence region.}
    \label{fig:beta_FWHM_eruption}
\end{figure}

The Lyman $\beta$ line width observed by SPICE during the eruption is shown in Figure~\ref{fig:beta_FWHM_eruption}. Values that are narrower than $\text{L}_\text{inst}$ have also been changed to the value of $\text{L}_\text{inst}$. The two white rectangles represent the prominence region and the disk region. The average line width in the prominence region is 0.09~nm and the average line width in the disk region is 0.09~nm. After accounting for the instrumental broadening, the intrinsic line width of the two regions is 0.05~nm. 

% \added{In \cite{heinzel2001soho}, the line widths of Lyman $\beta$ and Lyman $\gamma$ are around 0.05-0.06~nm (The instrumental broadening of Lyman lines was neglected) for the quiescent prominences observed on May 28 and June 2, 1999, by SUMER. This value is much less than our results. This suggests that the plasma conditions in eruptive prominences and quiescent prominences can be quite different, leading to different line broadening effects.}

In \cite{heinzel2001soho}, the line widths of Lyman~$\beta$ and Lyman~$\gamma$ are around 0.05--0.06~nm (the instrumental broadening of Lyman lines was neglected) for the quiescent prominences observed on May 28 and June 2, 1999, by SUMER. In \cite{tian2009hydrogen} , the line width of Lyman~$\beta$ is also around 0.05~nm. These values in \cite{heinzel2001soho} and \cite{tian2009hydrogen} are much less than our results. This suggests that the plasma conditions in quiescent prominences that are close to eruption can be quite different from those that are not, leading to different line-broadening effects. However, this possibility cannot be confirmed solely from the present data. The discrepancy may also be influenced by the remaining relative uncertainties in the instrumental profiles of SPICE compared with other instruments such as SUMER on SOHO. \added{Future modelling would allow us to determine whether there are narrower profiles in the eruption phase.}

\subsection{Line Profile}\label{profile}

\begin{figure*}
    \centering
\includegraphics[width=\textwidth]{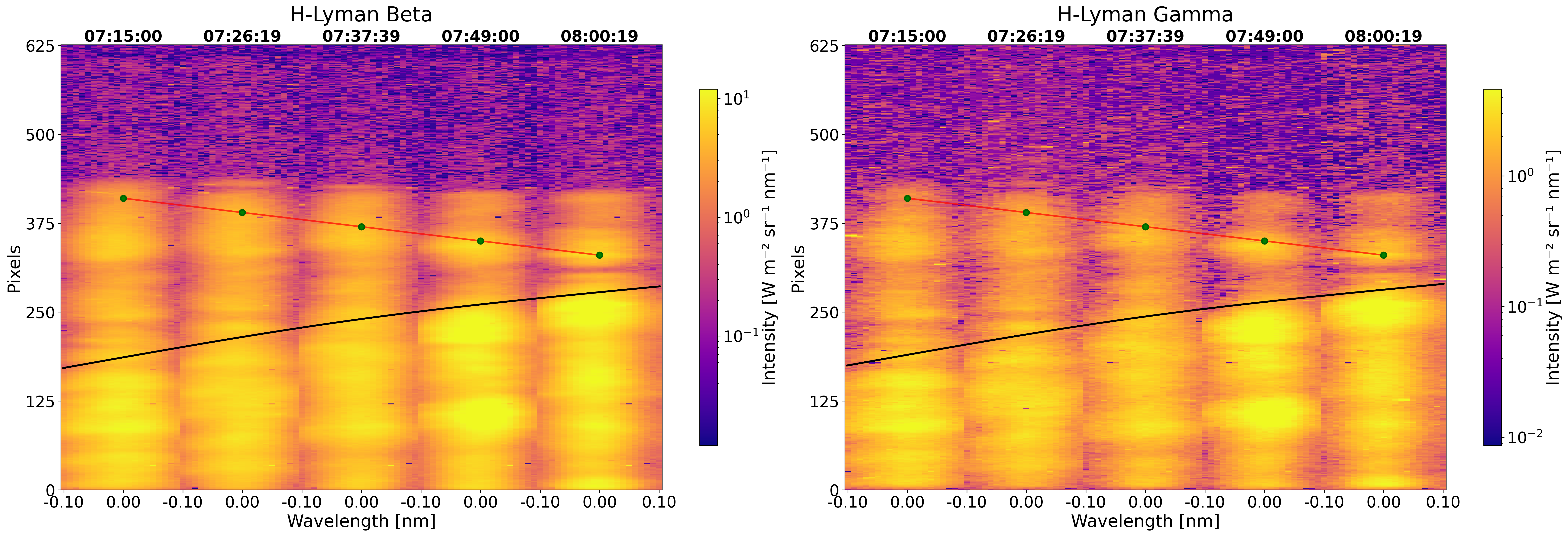}
    \caption{The Lyman $\beta$ and Lyman $\gamma$ lines spectra before the eruption from image of the detector for the exposures at the raster position x=5, 10, 15, 20, 25 in Figure~\ref{fig:beta_gamma} during pre-eruption phase. The center wavelength is 102.57~nm and 97.25~nm. The green dots and red lines refer to the positions of line profiles in Figure~\ref{fig:spectral_profiles}. The black curve is the boundary of the solar disc.}
    \label{fig:SPICE_slit_example}
\end{figure*}

\begin{figure}
    \centering
\includegraphics[width=0.5\textwidth]{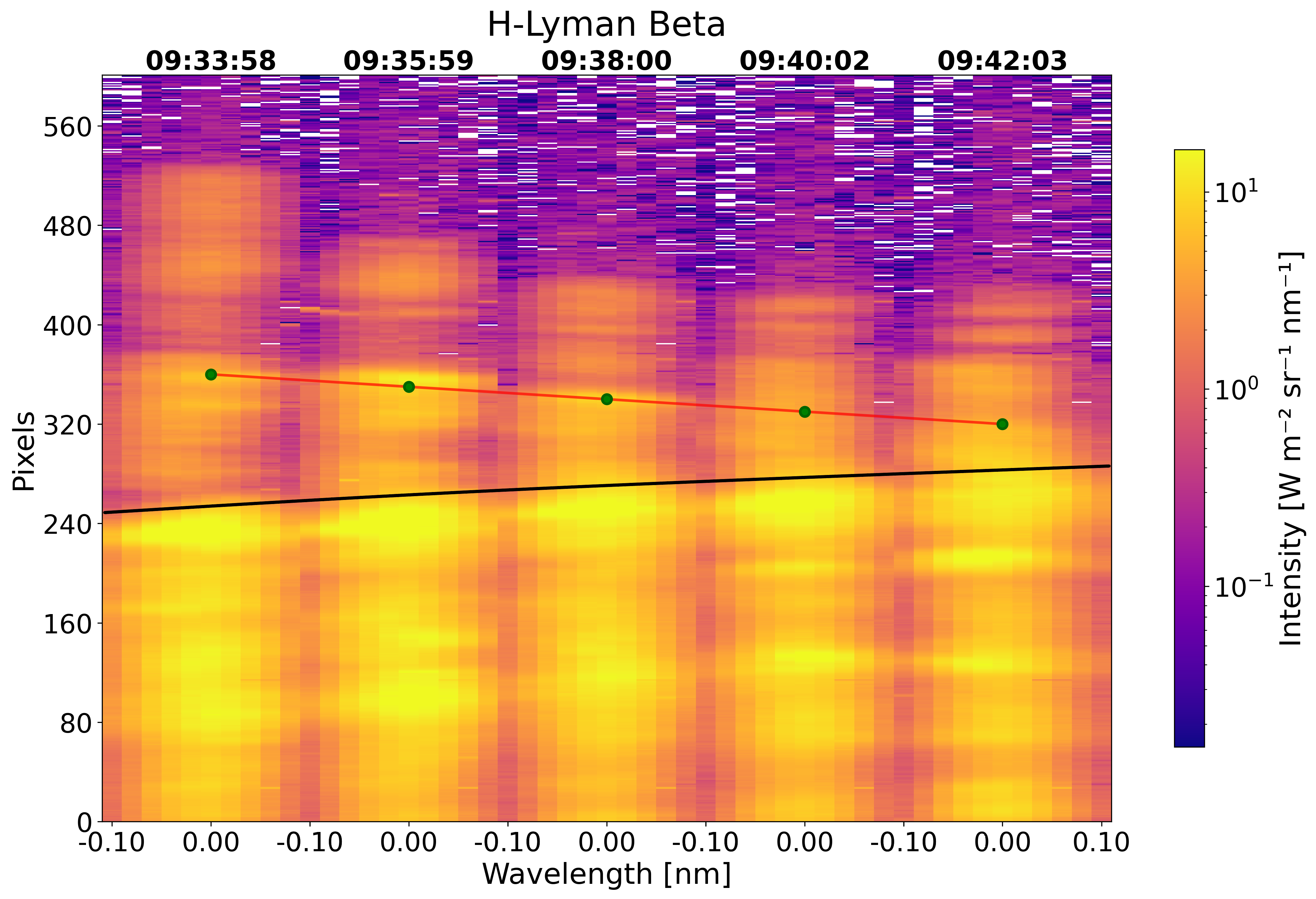}
    \caption{The Lyman $\beta$ lines spectra during eruption from image of the detector for the exposures at the raster position x=80, 82, 84, 86, 88 in Figure~\ref{fig:beta_eruption} during eruption phase. The center wavelength is 102.56~nm. The green dots and the red line refer to the positions of line profiles in Figure~\ref{fig:spectral_profiles_eruption}. The black curve is the boundary of the solar disc.}
    \label{fig:beta_slit_eruption}
\end{figure}

We present the spatial and temporal evolution of Lyman $\beta$ and Lyman $\gamma$ lines profiles in this section, which might provide information for diagnostics of the plasma conditions. Figure~\ref{fig:SPICE_slit_example} shows Lyman $\beta$ and Lyman $\gamma$ lines spectra before the eruption from image of the detector for the exposures at the raster position x=5, 10, 15, 20, 25 in Figure~\ref{fig:beta_gamma}. The Lyman $\beta$ line spectra during eruption from images of the detector for the exposures at the raster position x=80, 82, 84, 86, 88 are shown in Figure~\ref{fig:beta_slit_eruption}. The differences between the prominence region and disc region in Figure~\ref{fig:beta_slit_eruption} seem to be more pronounced than those in Figure~\ref{fig:SPICE_slit_example}. 

\begin{figure*}
    \centering
\includegraphics[width=\textwidth]{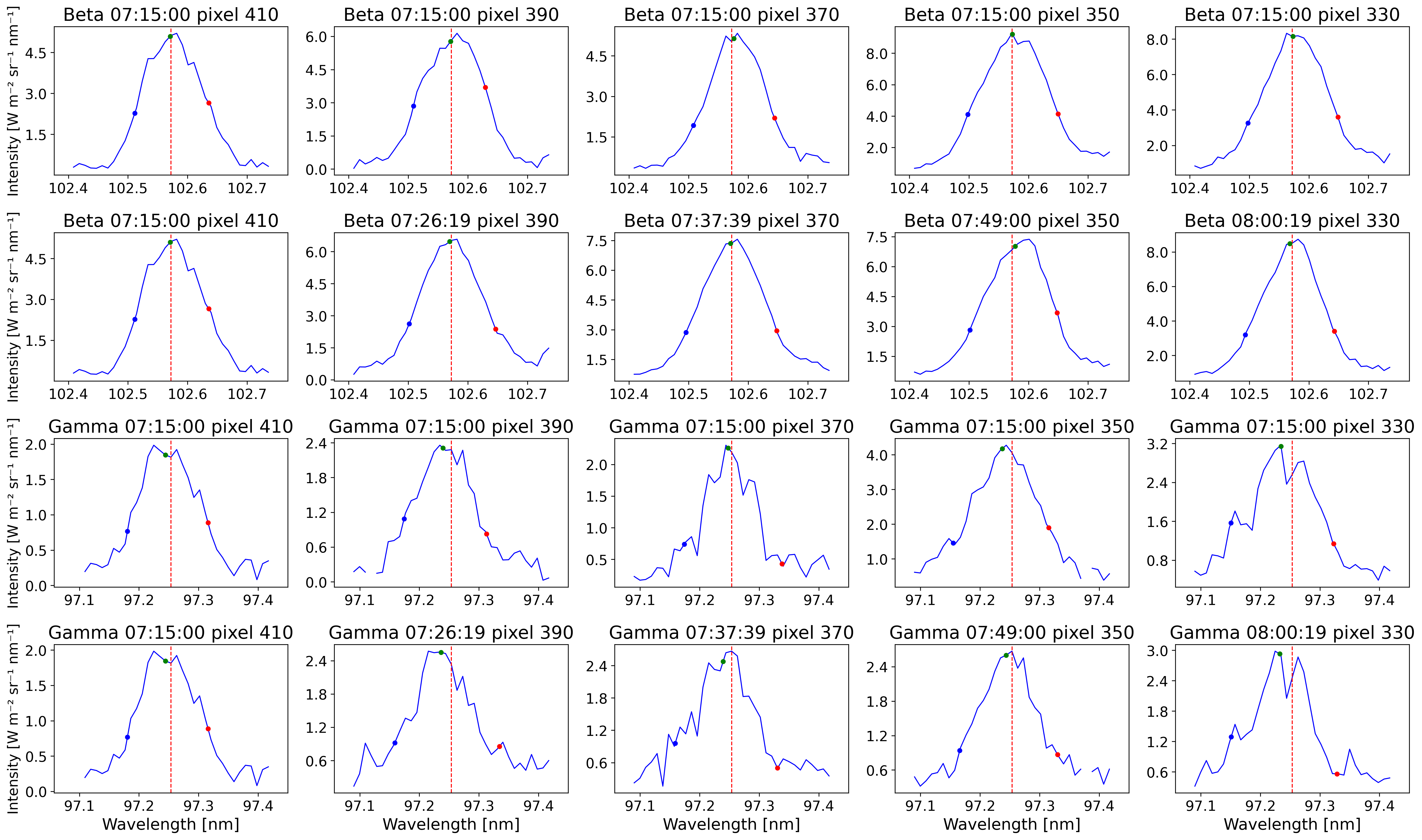}
    \caption{The Lyman $\beta$ and Lyman $\gamma$ lines profiles in the prominence region from SPICE observation before the eruption. The first and second rows are the Lyman $\beta$ line profiles. The first row is pixel 410, 390, 370, 350 and 330 at 07:15:00 UT in Figure~\ref{fig:SPICE_slit_example}. The second row is pixel 410, 390, 370, 350 and 330 from 07:15:00 UT to 08:00:19 UT, which is the same as the connection order of green points in Figure~\ref{fig:SPICE_slit_example}. The third and forth rows are the Lyman $\gamma$ line profiles. The time and location for these lines profiles are the same as the Lyman $\beta$ line. The blue point is the location CDF=0.12, the green point is the location CDF=0.50, and the red point is the location CDF=0.88. The red dashed lines are x=102.57~nm in the Lyman $\beta$ line profile and x=97.25~nm in the Lyman $\gamma$ line profile. These values are taken from the CDF=0.5 position of the average profile of the disk region (y-pixel 0-100 for all 30 }slits in Figure~\ref{fig:beta_gamma}).
    \label{fig:spectral_profiles}
\end{figure*}

Figure~\ref{fig:spectral_profiles} shows line profiles with details of spatial and temporal evolution before the eruption. The reference line centre (the red dashed lines) are taken from the CDF=0.5 position of the average profile of the disk region (y-pixel 0-100 for all 30 slits in Figure~\ref{fig:beta_gamma}). The center wavelength is 102.57~nm and 97.25~nm. The Lyman $\gamma$ line exhibits relatively more small-scale variations compared to Lyman $\beta$, although part of these features could be influenced by the difference in signal-to-noise ratio between the two lines. The temporal evolution is also more pronounced in the Lyman $\gamma$ line spectra. 

% \added{Figure~\ref{fig:spectral_profiles_eruption} shows Lyman $\beta$ line profiles with details of spatial and temporal evolution during the eruption. The reference line centre (the red dashed line) is taken from the CDF=0.5 position of the average profile of the disk region (the lower right white rectangle in Figure~\ref{fig:beta_FWHM_eruption}). The center wavelength is 102.56~nm. The pixels we choose in Figure~\ref{fig:spectral_profiles_eruption} generally show higher intensity compared with Lyman $\beta$ line profiles in Figure~\ref{fig:spectral_profiles}. We have calculated the Doppler shift of these line profiles. Most of the values are below the SPICE resolution limit (\sim 30~km~s$^{-1}$) \citep{plowman2025new}. We can conclude that no significant line of sight velocity can be detected.}

\begin{figure*}
    \centering
\includegraphics[width=\textwidth]{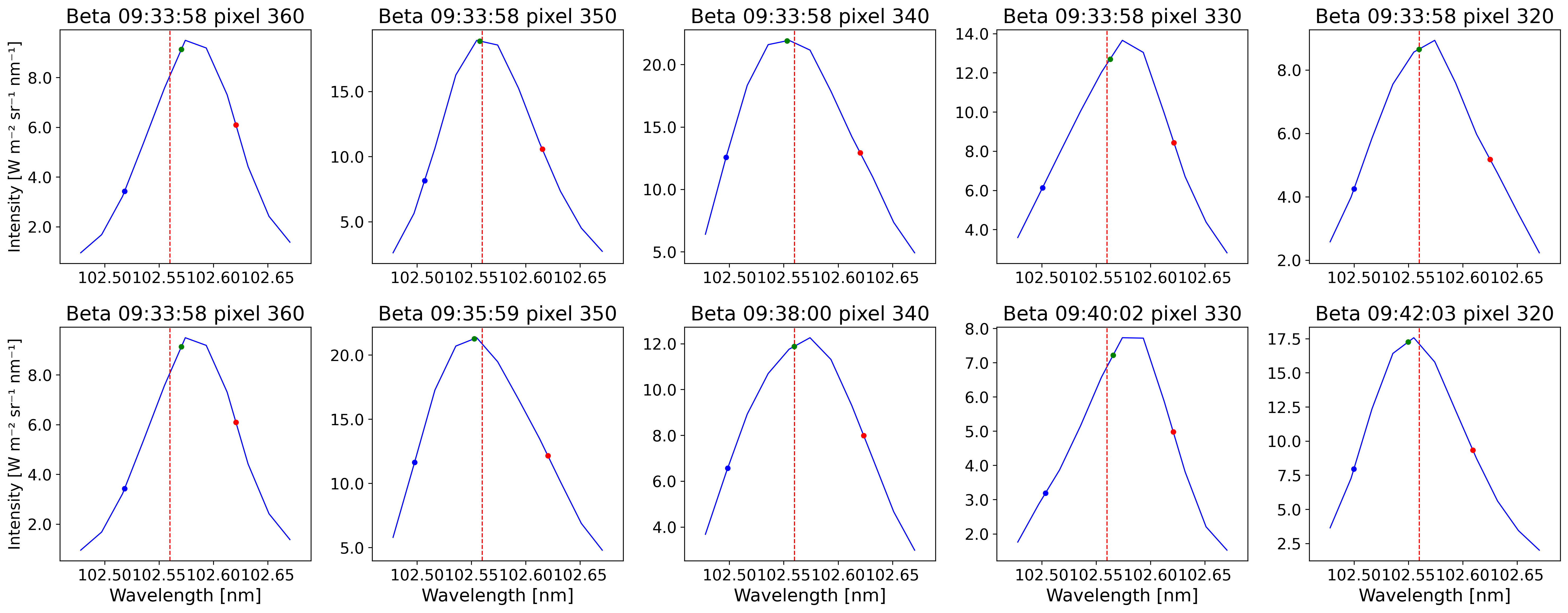}
    \caption{The Lyman $\beta$  lines profiles in the prominence region from SPICE observation during the eruption. The first row is pixel 360, 350, 340, 330 and 320 at 09:33:58 UT in Figure~\ref{fig:beta_slit_eruption}. The second row is pixel 360, 350, 340, 330 and 320 from 09:33:58 UT to 09:42:03 UT, which is the same as the connection order of green points in Figure~\ref{fig:beta_eruption_slit}. The blue point is the location CDF=0.12, the green point is the location CDF=0.50, and the red point is the location CDF=0.88. The red dashed lines are x=102.56~nm. The value is taken from the CDF=0.5 position of the average profile of the disk region (the lower right white rectangle in Figure~\ref{fig:beta_FWHM_eruption}).}
    \label{fig:spectral_profiles_eruption}
\end{figure*}

Figure~\ref{fig:spectral_profiles_eruption} shows Lyman $\beta$ line profiles with details of spatial and temporal evolution during the eruption. The reference line centre (the red dashed line) is taken from the CDF=0.5 position of the average profile of the disk region (the lower right white rectangle in Figure~\ref{fig:beta_FWHM_eruption}). The center wavelength is 102.56~nm. The pixels we choose in Figure~\ref{fig:spectral_profiles_eruption} generally show higher intensity compared with Lyman $\beta$ line profiles in Figure~\ref{fig:spectral_profiles}. We have calculated the Doppler shift of these line profiles. Most of the values are below the SPICE resolution limit (approximately 30~km~s$^{-1}$) \citep{plowman2025new}. We can conclude that no significant line-of-sight velocity can be detected.

Overall, the spectral profiles indicate that the eruption enhances spatial and temporal variations in line intensity and line-of-sight motions, reflecting dynamic changes in plasma conditions within the prominence.


% \begin{figure*}
%     \centering
% \includegraphics[width=\textwidth]{velocity_comparison.png}
%     \caption{\added{The velocity in the prominence region in slit x=5 and x=25 in Figure~\ref{fig:beta_gamma} calculated by the Lyman $\beta$ and Lyman $\gamma$ lines profiles. The reference profiles are from the disk region. }}
%     \label{fig:velocity_comparison}
% \end{figure*}

\section{Radial velocity estimation}\label{Sec: radial velocity}

Although Doppler shift of these line profiles is hard to determine, we still need to estimate the radial velocity by the observation as it is an important parameter in the Non-LTE modeling which we will present in the next paper and Lyman lines may be sensitive to Doppler dimming. We propose a new way of calculating the radial velocity from images. We can obtain the radial velocity at a certain time by 2 images of an instrument with the same point of view. This technique is fast and convenient only relying on a few assumptions.

\subsection{Direction of the eruption}

We need to derive the direction of the eruption to understand how the radial velocity is projected in the 2D image in Figure~\ref{fig:H_alpha}. According to \cite{mccauley2015prominence}, 75~\% of filament eruptions are radial. The rarity of non-radial solar eruptions can be attributed to the negative role of magnetic asymmetry in eruption initiation. Magnetic asymmetry inhibits, rather than facilitates, the eruption process. Numerical simulations by \cite{liu2024non} demonstrate that as the asymmetry of the magnetic flux distribution increases, the eruption direction deviates further from the radial path, while the eruption intensity weakens. In cases of strong asymmetry, eruptions may not occur at all, explaining why non-radial eruptions are less frequent than radial ones. Therefore, it is reasonable to assume radial eruption. In Figure~\ref{fig:H_alpha}, from the H$\alpha$ observation of El Teide at 10:07 UT, we identify the coordinates of two footpoints of this filament by choosing two points around the footpoints' area with similar intensity. Thus, we get the coordinates of the centre point of the filament and convert these to spherical coordinates.

\begin{figure}
    \centering
    \includegraphics[width=0.4\textwidth]{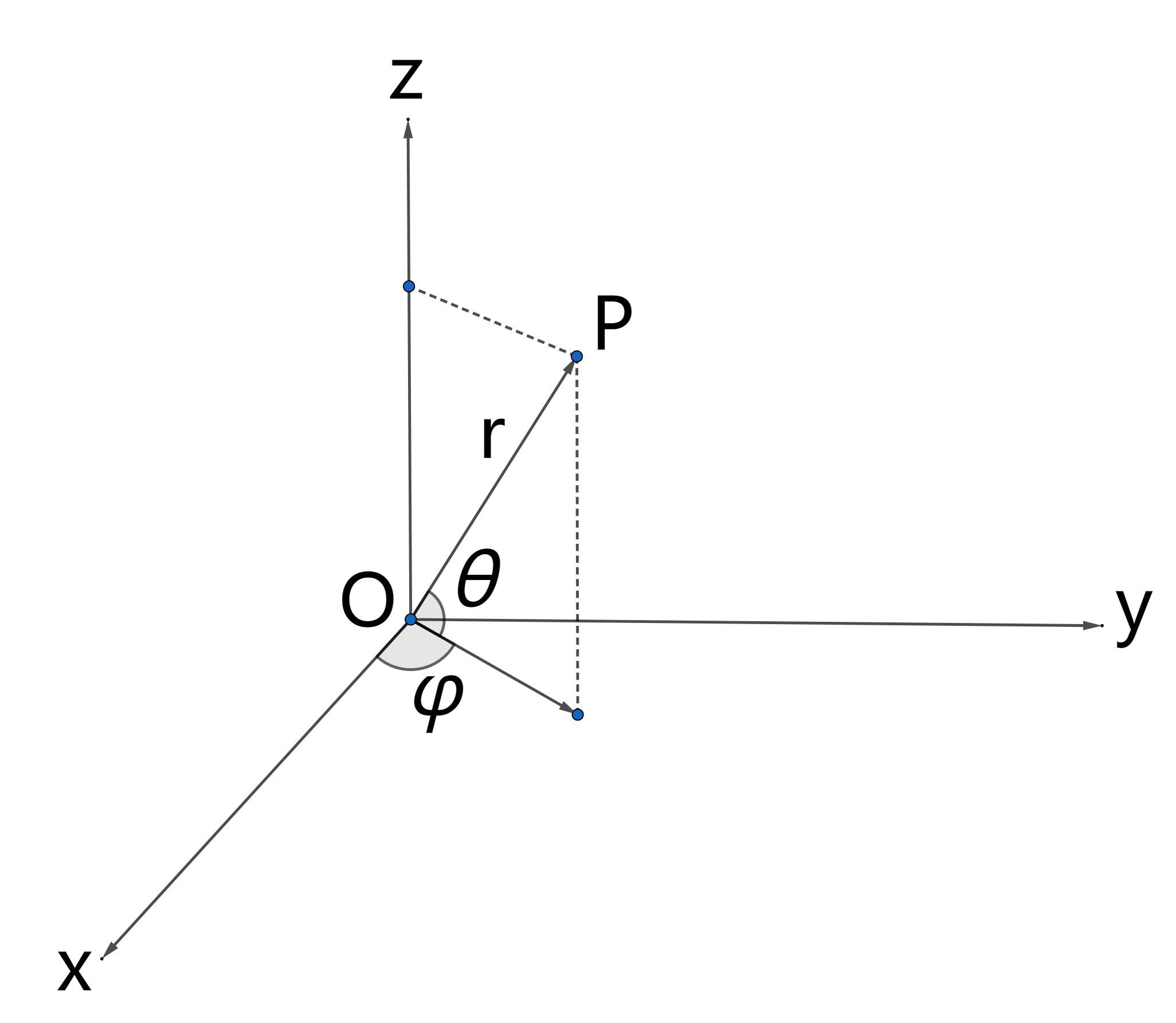}
    \caption{The illustration of the coordinates and angles}
    \label{fig:Coordinate}
\end{figure}

\begin{figure}
    \centering
    \includegraphics[width=0.4\textwidth]{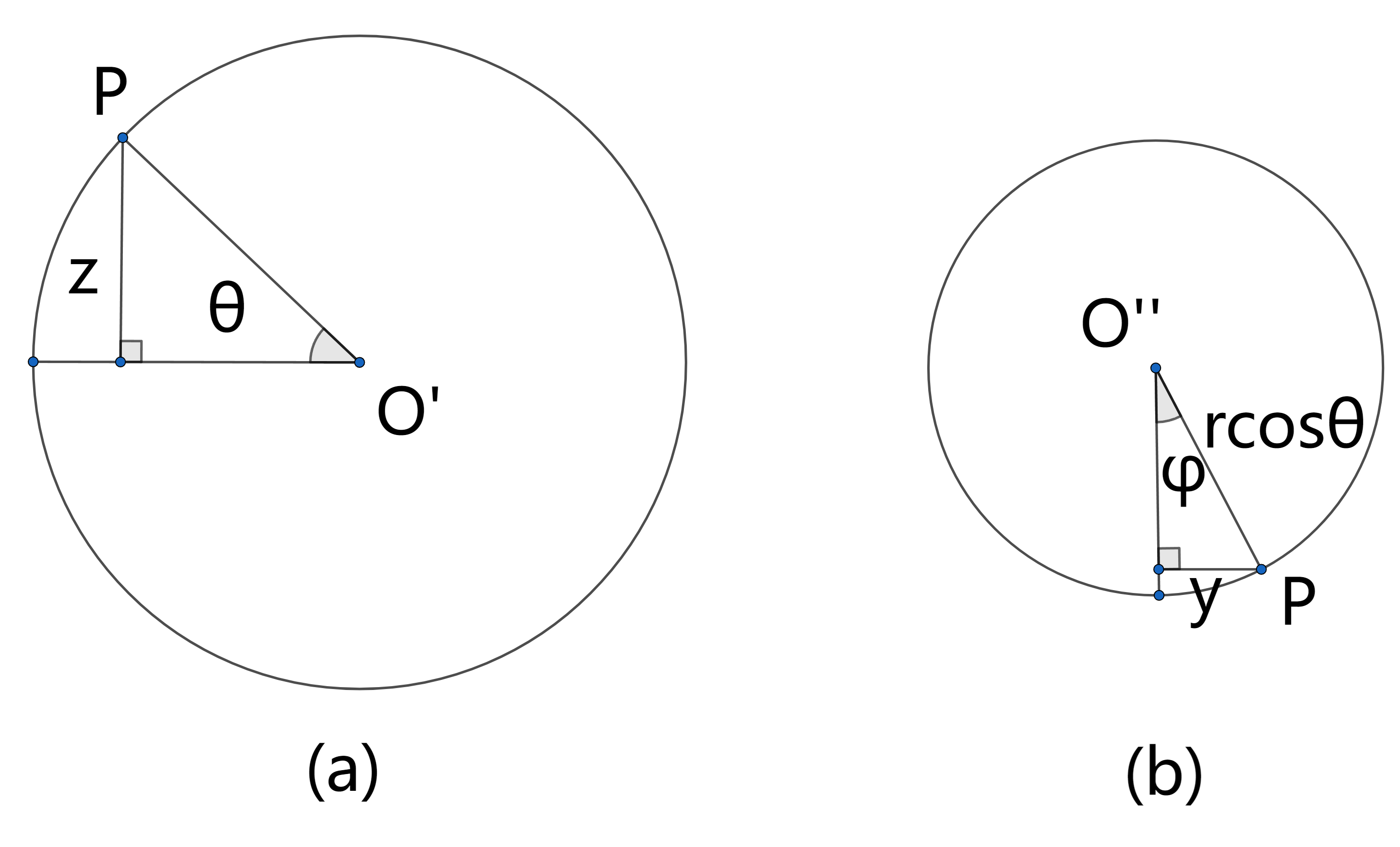}
    \caption{Two cross-sections of the sphere. The left one is through the dashed line (a) and the right one is through the dashed line (b) in Figure~\ref{fig:H_alpha}.}
    \label{fig:cross}
\end{figure}

\begin{figure*}
    \centering
    \includegraphics[width=0.9\textwidth]{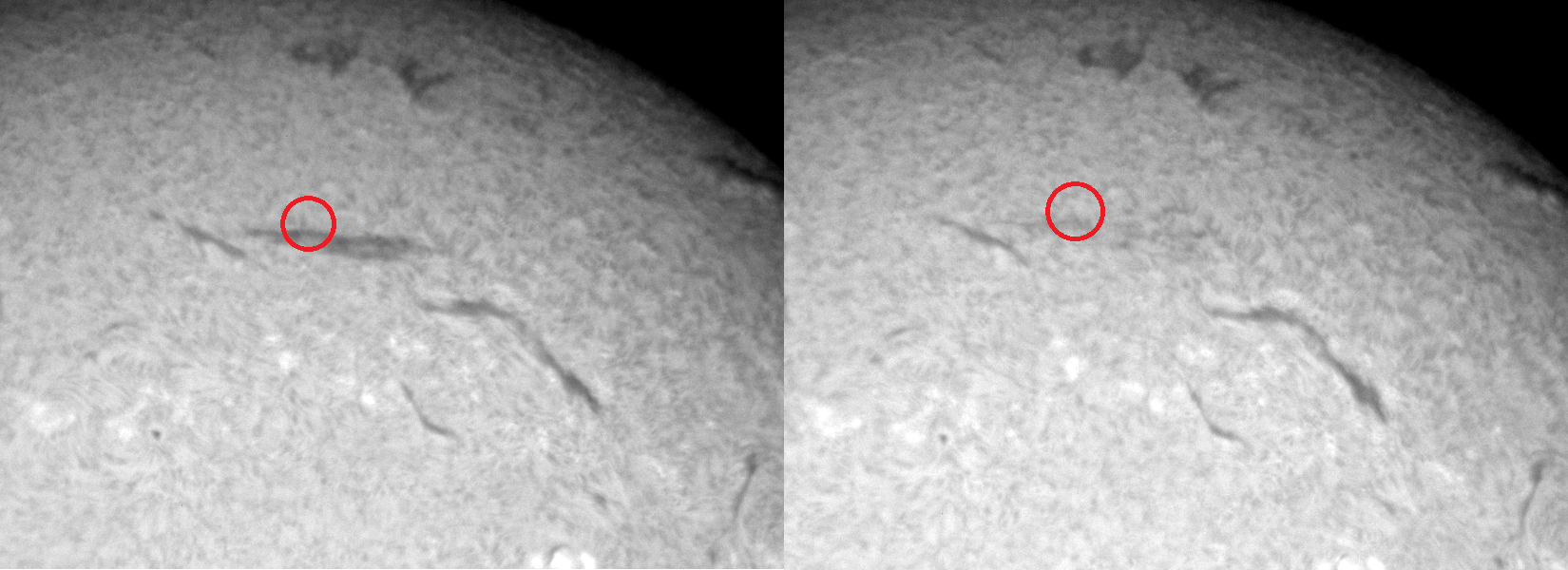}
    \caption{H$\alpha$ image of El Teide at 09:57 UT (left) and 10:17 UT (right), April 15, 2023. The filament is in the same region as Figure~\ref{fig:H_alpha}. The red circle shows the feature we track.}
    \label{fig:H_alpha_track}
\end{figure*}

We set the horizontal direction of the H$\alpha$ image as the $y$-axis, the vertical direction of the 2D image is the $z$-axis, and the $x$-axis is outward from the plane and perpendicular to the other axes. The $\theta$ and $\phi$ angles are shown in Figure~\ref{fig:Coordinate}. The coordinates of the center point of the filament footpoints (point P) projected in the $x=0$ plane are $(y, z)$. We find the relationship of $y$, $z$, $\theta$, and $\phi$ from two cross-sections of the sphere in Figure~\ref{fig:cross}. The left one is through the dashed line (a) and the right one is through the dashed line (b) in Figure~\ref{fig:H_alpha}. On the left side of Figure~\ref{fig:cross}, the cross-section contains the $z$-direction and point P. On the right side of Figure~\ref{fig:cross}, the cross-section is parallel to the $xy$-plane and contains point P. 

We assume the point \( P \) is on the solar surface. There should be some error about height in making this assumption, but it is negligible compared with the relatively large value of solar radius. After this step, we get the eruption direction by a vector \( \vec{OP} = (r, \theta, \phi) \) in the spherical coordinate. The point \( O \)  is the solar center.

So

\begin{eqnarray}
z &=& r \sin \theta \\
y &=& r \cos \theta \sin \phi
\end{eqnarray}

Then

\begin{eqnarray}
\theta &=& \arcsin\left(\frac{z}{r}\right) \\
\phi &=& \arcsin\left(\frac{y}{r \cos \theta}\right)
\end{eqnarray}

\subsection{Projection on 2D plane}
Assume the radial velocity is \( \vec{v} = (v_r, 0, 0) \) in spherical coordinates, or in Cartesian coordinates \( \vec{v} = (v_x, v_y, v_z) \):

\begin{eqnarray}
v_x &=& v_r \cos \theta \cos \phi \\
v_y &=& v_r \cos \theta \sin \phi \\
v_z &=& v_r \sin \theta
\end{eqnarray}

Also
\begin{eqnarray}
\vec{v} = \vec{v}_{x} + \vec{v}_{yz}
\end{eqnarray}
where the projection of the radial velocity on the \( y-z \) plane is:

\begin{align}
\vec{v}_{yz} &= \vec{v} - \vec{v}_{x} \notag \\
&= (v_r \cos \theta \cos \phi, v_r \cos \theta \sin \phi, v_r \sin \theta) \notag \\
&\quad - (v_r \cos \theta \cos \phi, 0, 0) \notag \\
&= (0, v_r \cos \theta \sin \phi, v_r \sin \theta)
\end{align}

So

\begin{eqnarray}
\left\lvert \vec{v}_{yz} \right\rvert \notag &=& v_r \sqrt{\left(\cos \theta \sin \phi\right)^2 + \left(\sin \theta\right)^2}
\end{eqnarray}

Since the velocity $v$ is assumed to be entirely in the radial direction

\begin{eqnarray}
\left\lvert \vec{v}_{yz} \right\rvert  &=& \left\lvert \vec{v} \right\rvert \sqrt{\left(\cos \theta \sin \phi\right)^2 + \left(\sin \theta\right)^2}
\label{rad}
\end{eqnarray}

\subsection{Calculation of radial velocity}
We assume  that the eruption is uniformly accelerated motion, and that different parts of the filament have the same speed during eruption. According to \cite{zhang2001temporal}, constant acceleration rate fitting is not adequate for some events. Nevertheless, our calculation is only based on 2 images with a 20~min interval, which is a relatively short period. Therefore, we assume the acceleration in the initiation phase of CME is stable. This assumption should be reasonable. Since the filament began to erupt at 09:57 UT, we used 09:57 UT and 10:17 UT images to identify the radial velocity at 10:07 UT.

Figure~\ref{fig:H_alpha_track} shows a feature that we can identify in the 09:57 UT and 10:17 UT images. After finding the coordinates of the circled features on these two images, the velocity of the filament eruption in the y-z plane can be calculated. Then, we can use the Equation~\ref{rad} to obtain the radial velocity. 10:07 UT is the last time we could identify the radial velocity by the H$\alpha$ observation of El Teide. This calculation shows that the radial velocity should be around $31 \pm 7$ km~s$^{-1}$ at 10:07 UT. The uncertainty of the result comes from the uncertainty of the location of the circled ends (about $\pm 7$~pixels), which is ambiguous in Figure~\ref{fig:H_alpha_track}. There are uncertainties from other sources: (1) if the radial eruption assumption is not satisfied, the derived velocity would have error depending on the inclined angle to the radial direction; (2) if the uniform acceleration assumption is not satisfied, the derived velocity would have an error depending on the pattern of acceleration process. 

\begin{figure*}
    \centering
    \includegraphics[width=\textwidth]{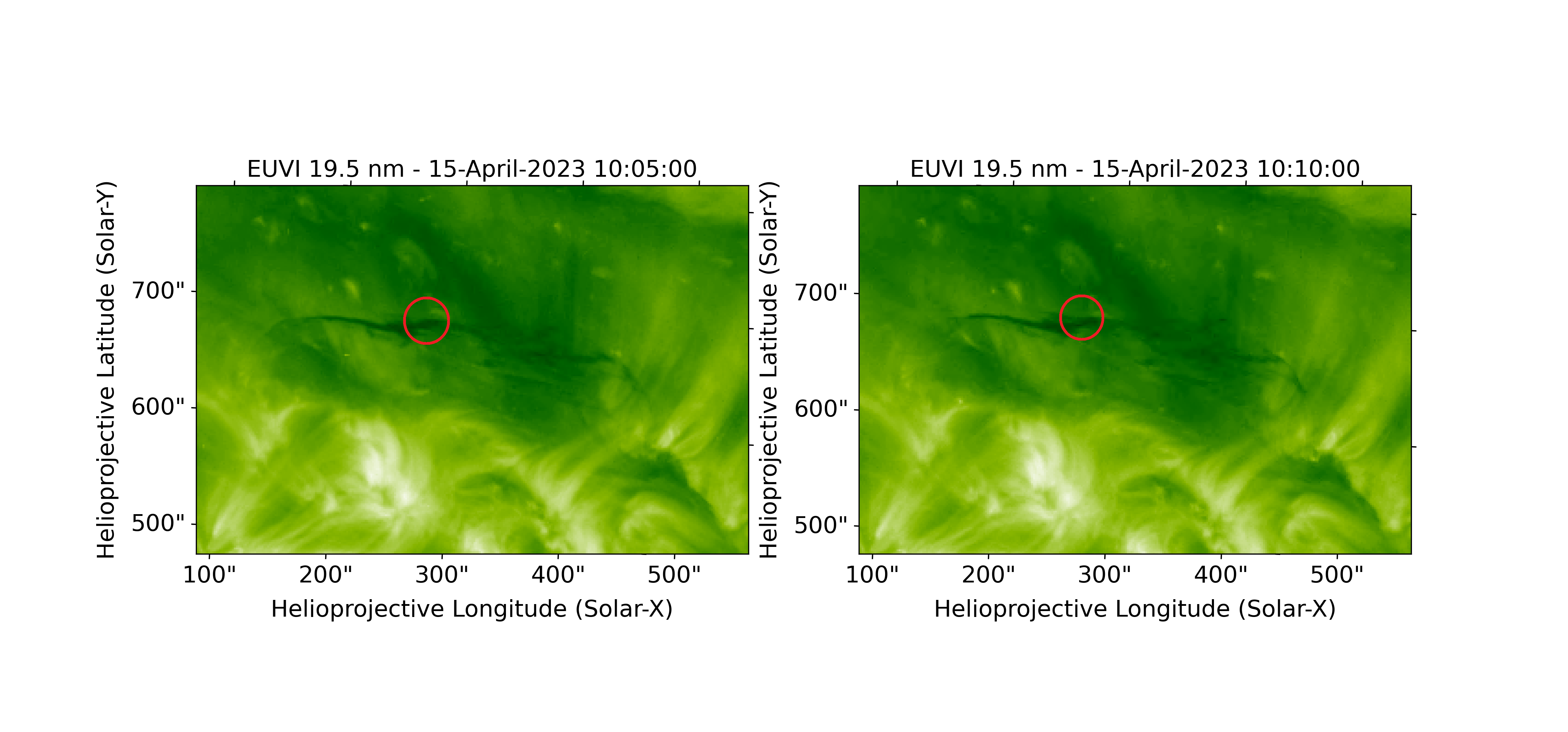}
    \caption{19.5~nm observation of EUVI-A at 10:05 UT (left) and 10:10 UT (right), April 15, 2023. The filament is in the same region as Figure~\ref{fig:H_alpha}. The red circle shows the feature we track.}
    \label{fig:EUVI_track}
\end{figure*}

We apply the same technique for the EUVI-A observation for this event as cross validation (Fig.~\ref{fig:EUVI_track}). This would be useful because STEREO is not on the Sun-Earth line on April 15, 2023. The result shows that the radial velocity should be around $44 \pm 5$ km~s$^{-1}$ at 10:07:30. The uncertainty of the result comes from the uncertainty of the location of the circled ends (about $\pm 5$~pixels). This result is quite close to our calculation using the H$\alpha$ observation of El Teide. The difference in the two results could be caused by the uncertainty of the location brought about by the two different observations at two different wavelengths. Nevertheless, we can see that results of calculation are relatively stable.  We set the result as the upper limit of radial velocity of the SPICE observation before 10:00 UT, April 15, 2023. Considering the radial velocity derived from these observations, only a modest Doppler shift on the hydrogen Lyman lines is expected. This is consistent from our previous conclusion that most of Doppler shift are below the resolution limit. For Lyman-$\beta$, the estimated Doppler shift ($\sim 0.015$~nm) is comparable to the line width, suggesting that the resonance scattering intensity could be reduced to some extent.

\section{Conclusions}\label{sec:conclusion}
In this study, we analyse the first dedicated observation of an off-limb eruptive prominence by the Spectral Imaging of the Coronal Environment (SPICE) which took place on April 15, 2023. 

We have analyzed the spatial variation of the integrated intensity of the Lyman $\beta$ and Lyman $\gamma$ lines. The solar disc exhibits high intensity due to its dense plasma and efficient collisional excitation, while the corona shows low intensity consistent with its low density. The ratio of Lyman $\beta$ to Lyman $\gamma$ remains roughly constant in the disc and prominence, suggesting similar radiative transfer conditions, but varies widely in the corona due to low signal levels. During the eruption, the prominence intensity of Lyman $\beta$ in region 1 increases substantially, occasionally surpassing the Lyman $\beta$ intensity in the disc, indicating enhanced heating, increased hydrogen column density, or compression. But eruptive prominence region 2 shows lower intensity and smaller line width compared with pre-eruption phase. This might be due to the low density in eruptive prominence region 2. 

We analyzed the line widths of the Lyman $\beta$ and Lyman $\gamma$ lines using the quantile method. In the pre-eruption phase, the prominence region exhibits similar intrinsic line widths (0.10~nm for Lyman $\beta$, 0.16~nm for Lyman $\gamma$) compared with the solar disk (0.09~nm for Lyman $\beta$, 0.13~nm for Lyman $\gamma$). During the eruption, the line widths of the prominence and disk regions become comparable (intrinsic width $\sim$ 0.05~nm). 

Compared with previous observations of quiescent prominences, our results show that eruptive prominences can display significant variations in line width, highlighting the strong influence of eruption processes on plasma dynamics and line broadening. But this possibility cannot be established solely from the present data. The difference may also be related to remaining relative uncertainties in the instrumental profiles of SPICE compared with other instruments such as SUMER on SOHO. We analyzed the spatial and temporal evolution of the Lyman $\beta$ and Lyman $\gamma$ line profiles to diagnose plasma conditions. Overall, the spectral profiles indicate that the eruption enhances spatial and temporal variations in line intensity and line-of-sight motions, reflecting dynamic changes in plasma conditions within the prominence.

We present a method to estimate the radial velocity through a 2D plane image. This offers an efficient way of calculating the radial velocity of an eruptive filament from a pair of 2D images. The upper limit of radial velocity of the SPICE observation is about $40$ km~s$^{-1}$before 10:00 UT, April 15, 2023. For Lyman-$\beta$, the estimated Doppler shift ($\sim 0.015$~nm) is comparable to the line width, suggesting that the resonance scattering intensity could be reduced to some extent. 

In future work, we will extend our analysis by applying Non-LTE radiative transfer modeling to reproduce the observed Lyman $\beta$ and Lyman $\gamma$ line profiles. This will allow us to constrain the plasma parameters of the prominence and refine our understanding of the physical processes governing its evolution.

\section*{Acknowledgements}
We would like to thank the teams involved in the development and operation of the EUI and SPICE instruments and data release. We are also grateful to the SunPy team for providing the tools used in our data analysis. The authors thank Lyndsay Fletcher for her comments on this paper. We also thank the referee for useful comments that resulted in a better paper. YZ is supported by the China Scholarship Council (No. 202206120056). NL acknowledges support from UK Research and Innovation's Science and Technology Facilities Council under grant award numbers ST/T000422/1 and ST/X000990/1. TAK was supported by Solar Orbiter/SPICE funding to NASA Goddard Space Flight Center.

Solar Orbiter is a space mission of international collaboration between ESA and NASA, operated by ESA. The EUI instrument was built by CSL, IAS, MPS, MSSL/UCL, PMOD/WRC, ROB, LCF/IO with funding from the Belgian Federal Science Policy Office (BELSPO/PRODEX PEA 4000112292 and 4000134088); the Centre National d’Etudes Spatiales (CNES); the UK Space Agency (UKSA); the Bundesministerium für Wirtschaft und Energie (BMWi) through the Deutsches Zentrum für Luft- und Raumfahrt (DLR); and the Swiss Space Office (SSO).
The development of SPICE has been funded by ESA member states and ESA. It was built and is operated by a multi-national consortium of research institutes supported by their respective funding agencies: STFC RAL (UKSA, hardware lead), IAS (CNES, operations lead), GSFC (NASA), MPS (DLR), PMOD/WRC (Swiss Space Office), SwRI (NASA), UiO (Norwegian Space Agency).
 Solar Orbiter is a space mission of international collaboration between ESA and NASA, operated by ESA. The EUI instrument was built by CSL, IAS, MPS, MSSL/UCL, PMOD/WRC, ROB, LCF/IO with funding from the Belgian Federal Science Policy Office (BELSPO/PRODEX PEA 4000112292 and 4000134088); the Centre National d’Etudes Spatiales (CNES); the UK Space Agency (UKSA); the Bundesministerium für Wirtschaft und Energie (BMWi) through the Deutsches Zentrum für Luft- und Raumfahrt (DLR); and the Swiss Space Office (SSO).

 STEREO is the third mission in NASA’s Solar Terrestrial Probes
(STP) program.  The STEREO/SECCHI instrument suite was produced by a consortium of NRL (US), LMSAL (US), NASA/GSFC (US), RAL (UK), UBHAM (UK), MPS (Germany), CSL (Belgium), IOTA (France), and IAS (France).

%%%%%%%%%%%%%%%%%%%%%%%%%%%%%%%%%%%%%%%%%%%%%%%%%%
\section*{Data Availability}

This research used version 5.0.0 of the SunPy open source software package. We obtained EUI and SPICE data through the ESA Solar Orbiter Archive. We have used EUI Data Release 6.0: (DOI: https://doi.org/10.24414/z818-4163) and SPICE Data Release 5.0: (DOI: https://doi.org/10.48326/idoc.medoc.spice.5.0).

%%%%%%%%%%%%%%%%%%%% REFERENCES %%%%%%%%%%%%%%%%%%

% The best way to enter references is to use BibTeX:

\bibliographystyle{mnras}
\bibliography{paper1} % if your bibtex file is called example.bib

% Alternatively you could enter them by hand, like this:
% This method is tedious and prone to error if you have lots of references
%\begin{thebibliography}{99}
%\bibitem[\protect\citeauthoryear{Author}{2012}]{Author2012}
%Author A.~N., 2013, Journal of Improbable Astronomy, 1, 1
%\bibitem[\protect\citeauthoryear{Others}{2013}]{Others2013}
%Others S., 2012, Journal of Interesting Stuff, 17, 198
%\end{thebibliography}

%%%%%%%%%%%%%%%%%%%%%%%%%%%%%%%%%%%%%%%%%%%%%%%%%%

%%%%%%%%%%%%%%%%% APPENDICES %%%%%%%%%%%%%%%%%%%%%

\appendix

%%%%%%%%%%%%%%%%%%%%%%%%%%%%%%%%%%%%%%%%%%%%%%%%%%

% Don't change these lines
\bsp	% typesetting comment
\label{lastpage}
\end{document}